\documentclass[trackchanges,twocolumn,tighten]{aastex701}

\makeatletter

\let\oldappendix\appendix
\renewcommand{\appendix}{%
  \oldappendix
  \@addtoreset{equation}{section}
}

\newcommand{\uatFix}[2]{\href{http://astrothesaurus.org/uat/#2}{#1 (#2)}}

\makeatother

\usepackage{amsmath}
\usepackage{mathtools}

\newcommand{\tn}[1]{\textnormal{#1}}
\newcommand{\lrp}[1]{\left(#1\right)}

\newcommand{\spiro}{\texttt{SPIRO}}     
\newcommand{\sgn}{\operatorname{sgn}}

\begin{document}


\title{From Internal Collision to Photon Escape:\\ First-Principles Modeling  of Radiation-Mediated Shocks in Gamma-Ray Burst Photospheres}
\shorttitle{From internal collision to photon escape}
\shortauthors{Nordin Nobuoka et al.}

\author[orcid=0009-0004-5674-4786,sname='Nordin Nobuoka']{Jona Nordin Nobuoka}
\affiliation{KTH Royal Institute of Technology, Department of Physics, SE-10691 Stockholm, Sweden}
\affiliation{The Oskar Klein Centre for Cosmoparticle Physics, AlbaNova University Centre, SE-10691 Stockholm, Sweden}
\email{}

\author[orcid=0000-0001-7414-5884,sname='Alamaa']{Filip Alamaa}
\affiliation{KTH Royal Institute of Technology, Department of Physics, SE-10691 Stockholm, Sweden}
\affiliation{The Oskar Klein Centre for Cosmoparticle Physics, AlbaNova University Centre, SE-10691 Stockholm, Sweden}
\email{}

\author[orcid=0000-0002-9769-8016,sname='Ryde']{Felix Ryde}
\affiliation{KTH Royal Institute of Technology, Department of Physics, SE-10691 Stockholm, Sweden}
\affiliation{The Oskar Klein Centre for Cosmoparticle Physics, AlbaNova University Centre, SE-10691 Stockholm, Sweden}
\email{}

\author[orcid=0000-0002-0642-1055,sname='Lundman']{Christoffer Lundman}
\affiliation{KTH Royal Institute of Technology, Department of Physics, SE-10691 Stockholm, Sweden}
\affiliation{The Oskar Klein Centre for Cosmoparticle Physics, AlbaNova University Centre, SE-10691 Stockholm, Sweden}
\email{}

\begin{abstract}
Modeling subphotospheric shocks in a gamma-ray burst (GRB) is challenging due to the various timescales that must be resolved, and the fact that the same radiation dynamically mediates the shocks while forming the observed signal.
Here, we present the first self-consistent radiation-hydrodynamic simulation of a subphotospheric internal collision, following the system from formation and propagation of forward and reverse radiation-mediated shocks all the way to photon decoupling and free streaming toward the observer. The simulation evolves the plasma and photon field with full Compton coupling, including the feedback on the hydrodynamic flow. As the ejecta expands and the optical depth decreases, both shocks broaden and the radiation field becomes highly non-thermal. Surprisingly, we find that the reverse shock remains completely radiation-mediated down to upstream optical depths of order a few $\times 10^{-1}$, which indicates that Compton coupling is important even in moderately optically thin regions.
The photons undergo last scattering over a broad range of radii rather than at a single photospheric surface. The light curve shows a late, quasi-thermal post-cursor produced by photons that decouple upstream of the reverse shock, which could be searched for in observations. The emitted time-integrated spectrum is GRB-like, with a low-energy photon index $\alpha \sim -1$ and a high-energy photon index $\beta \sim -2.5$. These results show how radiation-mediated shocks evolve close to the photosphere and how they shape the emitted photon field.
\end{abstract}


\section{Introduction}

Gamma-ray burst (GRB) prompt emission has been studied for more than five decades, yet its physical origin remains unsettled.
In optically-thin models, dissipation occurs above the photosphere and the radiation is due to a non-thermal process, such as synchrotron emission from particles energized in collisionless internal shocks or magnetic reconnection \citep[e.g., ][]{ReesMeszaros1994, DaigneMochkovitch1998, SpruitDaigneDrenkhahn2001}. An alternative possibility is that a substantial fraction of the prompt emission is formed near or below the photosphere, where the jet remains optically thick and radiation is dynamically important \citep[e.g., ][]{Paczynski1986, ReesMeszaros2005}. This possibility is attractive because the photospheric region is unavoidable in relativistic outflows \citep[][]{ShemiPiran1990, DaigneMochkovitch2002} and because many observed GRB spectra appear thermal \citep[e.g., ][]{Ryde2004, Ryde2010, Chen2024}.

While some spectra appear thermal, others require dissipation acting on a pre-existing radiation field rather than emission from a purely passive fireball \citep[e.g., ][]{Iyyani2015, SamuelssonRyde2023}. One of the most promising dissipation mechanisms in the optically-thick family is radiation-mediated shocks \citep[RMSs, see][for a review]{LevinsonNakar2020}. Recollimation shocks appear to be a generic feature of multidimensional jet simulations and are expected to be radiation-mediated in GRBs \citep{Lazzati2009, Lopez-Camara2013, Ito2015, Gottlieb2019, Pais2024}. Likewise, many internal collisions in GRB jets should occur below the photosphere, and therefore produce RMSs rather than collisionless shocks \citep{Bromberg2011b}. Finally, recent fits of RMS-based models to prompt-emission data provide further support for the relevance of these shocks to observations \citep{Samuelsson2022, Wistemar2025b, Wistemar2025a}. However, whether or not RMSs can self-consistently change the jet radiation field to form the observed spectrum and observed temporal properties has not yet been fully addressed. 

That question has proven difficult to answer. Subphotospheric dissipation occurs in an optically-thick, radiation-mediated regime where the plasma is tightly coupled to the photons; the same photons that are eventually observed also mediate the shock, shaping the velocity and pressure structure of the plasma. Moreover, close to the photosphere when the scattering time becomes comparable to the local dynamical time, the photon mean free path becomes macroscopic. At this point the system becomes intrinsically time dependent, since the RMSs will not be able to adjust to the changing plasma properties. As a result of these complexities, previous work has relied on a variety of approximations, simplifying various parts of the full problem \citep[although see][in the case of GRB 170817]{LundmanBeloborodov2021}.

Typical approximations include steady-state, planar RMS calculations, which have been useful to study the microphysics of RMSs and their spectra \citep{Ito2018, Lundman2018}. However, by construction such simulations neither follow the expansion of the outflow, nor the release of radiation through the photosphere. Another approach is to perform multidimensional hydrodynamic simulations with Monte Carlo radiation treated in post-processing \citep{Ito2015, Lazzati2016, ParsotanLazzati2018}. This approach captures the effects of geometric complexity and higher dimensional behaviors such as large-scale turbulence. However, it cannot recover the detailed shock structure and the correct hydrodynamical profile close to the photosphere since the radiation is not allowed to react back on the plasma. The shock width is instead set by the grid size; the radiation will not get the proper energy and spectral shape from such a shock. Finally, \citet{Samuelsson2022} introduced the Kompaneets RMS approximation (KRA) by noting the similarity between the energy dissipation process in an RMS and thermal Comptonization. Due to its low computational cost, the KRA can be used to fit GRB data \citep{Samuelsson2022, Wistemar2025b, Wistemar2025a}. However, it was designed for optically thick shocks and cannot capture the dynamics close to the photosphere.

In this paper, we effectively remove the above approximations by introducing {\tt {SPIRO}}, a relativistic radiation-hydrodynamic code that couples hydrodynamics to Monte Carlo radiative transfer, following Compton scattering, photon transport, and radiation feedback on the plasma self-consistently. We study a deliberately simple setup with minimal ingredients, namely a spherically symmetric internal collision below the GRB photosphere. While simple, this approximation is useful because it isolates the essential ingredients of a subphotospheric shock without introducing unnecessary model complexity. The code captures the initial compression, the formation and propagation of forward and reverse RMSs, the weakening of the shocks at low optical depths, the possible formation of collisionless subshocks, and finally, the release of radiation to the observer.

We find that as the reverse and forward shocks propagate outward, their radiation fields become increasingly non-thermal. The plasma Compton coupling remains strong even in regions where the local optical depth is significantly below unity. When the photons decouple, they do so over a broad range of radii. The predicted observations consists of a structured light curve pulse and a broad GRB-like spectrum. The model connects the internal dynamics of subphotospheric shocks directly to observables and shows that a dissipation process with only a few ingredients can account for key properties of the prompt emission.

The paper is structured as follows. In Section \ref{sec:SPIRO}, we introduce the numerical code \texttt{SPIRO} and explain its components. A reader that is mainly interested in GRB-related results can skip this section. The initial conditions for the subphotospheric internal collision are given in Section \ref{sec:initial_conditions}. The results are presented in Section \ref{sec:results} and a discussion with an emphasis on model assumptions is given in Section \ref{sec:discussion}. Our conclusions are summarized in Section \ref{sec:conclusion}.

\section{SPIRO}\label{sec:SPIRO}

In this section, we introduce \spiro, which is a 1D special relativistic hydrodynamical code with coupled Monte Carlo photons. In short, the code works by evolving the hydrodynamics for a short time, $\delta t$, using the open source special relativistic hydrodynamical code \texttt{GAMMA} \citep{Ayache2022}. Then, Monte Carlo photons are propagated through the hydro-grid for the same duration $\delta t$, where they can interact with the plasma via Compton scattering.
Every change in energy and momentum of the photons due to scattering is absorbed into the plasma and is accounted for in the next hydro-step.

\subsection{Hydrodynamics}

\verb|SPIRO| is built on top of the open source special relativistic hydrodynamics code \verb|GAMMA| \citep[][see also \citet{Ayache2020}]{Ayache2022}. With its Lagrangian-Eulerian approach, \texttt{GAMMA} is well suited to model astrophysical flows, including GRBs during both the prompt and the afterglow phase.

Special relativistic effects are fully accounted for but the effects of gravity or magnetic fields are not modeled. We use a non-relativistic monoatomic ideal equation of state, with adiabatic index $5/3$. This is accurate so long as the comoving plasma temperature is low enough for the protons to be non-relativistic. Note that the effective adiabatic index of the outflow is completely determined by the photons in the optically thick regime.

The simulation domain is consists of \textit{cells} delimited by \textit{interfaces}. The flux of energy, mass, and momentum across each interface are computed using the approximate Riemann solver HLLC \citep{MignoneBodo2006}. \texttt{GAMMA} can be used for simulations in 1D or 2D, however, \texttt{SPIRO} in its present version is limited to 1D plane parallel or spherically symmetric geometries.

\texttt{GAMMA} supports adaptive mesh refinement, which makes it possible to satisfy the local resolution requirements at each timestep using fewer cells. \texttt{SPIRO} uses circular regrid functionality in \texttt{GAMMA}, which means that every time a cell merges with its neighbor, another cell is split in half, keeping the total number of cells constant.

\subsection{Monte Carlo radiation}

The microphysics of an RMS occur at the length scale of the photon mean free path. At this scale, the radiation stops being collisionally bound to the plasma, which makes the fluid approximation implicitly assumed in a hydrodynamic simulation break down.
\verb|SPIRO| instead represents the radiation with Monte Carlo \textit{photon packets}.\footnote{The Monte Carlo photon packets will also be referred to as Monte Carlo photons or simply photons throughout the text.} A packet is characterized by its dimensionless energy, $\varepsilon \equiv E/m_e c^2$, where $E$ is the lab frame photon energy and $m_e$ is the electron mass, its direction, $\mu$, its radial position, $r$, and its weight, $w$. The direction parameter $\mu$ is defined as $\mu \equiv \cos \theta$, where $\theta$ is the lab frame angle between the local radial direction and the momentum vector of the photon. The weight $w$ denotes how many physical photons the packet represents; however, each photon packet moves and interacts as if it were a single photon. The method of modeling the radiation with Monte Carlo photon packets has been used in many previous works \citep{PeerWaxman2005, Beloborodov2011, Ito2015, Lazzati2016, ParsotanLazzati2018, Lundman2018}. 

We assume that the photons are unpolarized. Due to their high energies, diffraction effects will be negligible. Thus, the photons must propagate along straight lines between interactions. In a spherical geometry, as considered in this paper, the position and direction of a freely propagating photon packet evolves according to
\begin{equation}
    \label{eq:r_of_t}
    r(t) = \sqrt{r_0^2 + 2 r_0 \mu_0 c\Delta t +  (c\Delta t)^2}
\end{equation}
and
\begin{equation}
    \mu(t) = \frac{\mu_0 r_0 + c\Delta t}{\sqrt{r_0^2 + 2 r_0 \mu_0 c\Delta t +  (c\Delta t)^2}},
    \label{eq:mu_of_t}
\end{equation}
where $\Delta t \equiv t-t_0$, $r_0\equiv r(t_0)$, and $\mu_0\equiv\mu(t_0)$, $t$ is the time measured in the lab frame, and $t_0$ is a reference time (after which the photon has been travelling in a straight line).

An RMS that occurs in an unmagnetized GRB is photon rich, which means that the upstream photon-to-lepton ratio is high enough such that photon production in the shock transition region can be ignored \citep{Bromberg2011b}. However, if the GRB is moderately magnetized, photon generation via synchrotron may be substantial \citep{LundmanBeloborodov2019}. In this paper, we do not include photon production, i.e., we assume a photon rich outflow with negligible magnetization.

\subsection{Compton scattering}
In a fully ionized, unmagnetized, photon rich outflow, the spectral evolution is dominated by Compton scattering. 
This is also the microphysical process that mediates an RMS. For this reason, the only plasma-photon interaction that \verb|SPIRO| models is Klein-Nishina corrected Compton scattering.

The lab frame mean free time between Compton scatterings for a photon traveling through a thermal plasma is given by
\begin{equation}\label{eq:t_mf}
    t_\tn{mf} = \frac{1}{(1 - \beta_r\mu) n_e c \tilde\sigma(\varepsilon^{\prime}, \Theta_\tn{pl})},
\end{equation}
where $\beta_r$ is the lab frame dimensionless radial bulk velocity of the plasma, $n_e$ is the lab frame electron density, and $\tilde\sigma$ is the effective cross section given in Equation \eqref{eq:effective_sigma}, which depends on the dimensionless (comoving) plasma temperature $\Theta_\tn{pl}\equiv k_\tn{B} T_\tn{pl} / m_ec^2$, and comoving photon energy $\varepsilon^{\prime}$. We use the full Klein-Nishina cross section and assume that the electrons are in a thermal Maxwell-J\"uttner distribution.\footnote{To assume that the electrons are always thermal is a very good approximation in an optically thick GRB jet. The high photon-to-electron ratio means that the electron cooling time-scale is much shorter than $t_\tn{mf}$ \citep{Lundman2018}.}

The outcome of a scattering event depends on the initial momentum of the electron that the photon interacts with. 
However, the electron momentum cannot be sampled directly from the Maxwell-J\"uttner distribution. This is because the interaction rate depends on the Galilean relative velocity between the photon and the electron. Another source of interaction rate bias is the fact that the Klein-Nishina cross section depends on what energy the photon will have in the rest frame of the electron: the interaction probability will be suppressed if the photon has an energy comparable to $m_e c^2$ in the electron rest frame. More details are given in Appendix \ref{sec:kn_corrections}.

A Compton-scattering event between a photon packet and an electron in a cell is computed as follows. First an electron velocity is sampled based on the comoving temperature of the plasma in the cell, accounting for the Galilean relative velocity and the Klein-Nishina corrections to the cross section as explained above. The energy and direction of the photon packet is then Lorentz-transformed to the rest frame of the sampled electron. A scattering is performed in that frame and the new energy and direction for the photon packet is sampled using the Klein-Nishina formula. The photon is then transformed back to the lab frame. The change in energy and momentum of the photon (positive or negative), scaled by the weight of the packet, is then added to the plasma in the cell.

Pair production and annihilation becomes relevant when the comoving photon energies approach the electron rest mass \citep{LevinsonBromberg2008, Budnik2010, Katz2010, Ito2018, Lundman2018}.
Currently, \verb|SPIRO| is designed to work in the regime where pair production is negligible. To ensure this is indeed the case, \verb|SPIRO| dynamically computes local pair production rates in all cells. Pairs were found to be irrelevant for the two simulations mentioned in this paper, see Appendix \ref{app:pair_check} for details.

\subsection{Packet propagation}

The scattering probability is governed by an exponential distribution, which has a probability density function as 
\begin{equation}\label{eq:exponential_distr}
    p(t, t_{\rm mf}) = \frac{1}{t_\tn{mf}} \exp\left(-\frac{t}{t_\tn{mf}}\right),
\end{equation}
provided that $t_\tn{mf}$ does not change during the time $t$. Any scattering event will cause an instantaneous change in $t_\tn{mf}$, both by altering $\varepsilon$ and $\mu$ of the packet, as well as $\beta_r$ and $\Theta_\tn{pl}$ of the plasma in the cell that it currently inhabits. If $t_\tn{mf}$ only changes through instantaneous events, then we can construct a stochastically correct algorithm for packet propagation without needing to integrate $t_\tn{mf}$ along the photon path.

There are three additional ways that $t_\tn{mf}$, given in Equation \eqref{eq:t_mf}, can be altered during the propagation:
\begin{enumerate}
    \item The plasma properties change over time due to hydrodynamical evolution.
    \item There is some spatial variation in the plasma properties along the path of the packet.
    \item The packet travels for so long that the change in $\mu$ due to Equation \eqref{eq:mu_of_t} becomes significant. This is only relevant in a spherical geometry.
\end{enumerate}
If the temporal and spatial resolution is high enough, then the first two points can be handled by recalculating $t_\tn{mf}$ after every hydro-step and whenever the packet switches cell. These events will be referred to as \textit{hydro-evolution events} and \textit{cell change events}, respectively. The kinematic drift of $\mu$ is most relevant when $r \gtrsim ct_\tn{mf}$, i.e., close to or above the transparency radius. It can be dealt with by breaking up long propagation steps with \textit{kinematic drift events} that recalculate $t_\tn{mf}$ before it has time to change significantly. More details regarding the propagation are given in Appendix \ref{sec:prop_details}.

\verb|SPIRO|'s algorithm for photon propagation is as follows. For a photon packet at $r_0$ with direction $\mu_0$ at time $t_0$, the local value of $t_\tn{mf}$ is calculated and used to sample a candidate duration $\Delta t_\tn{SC}$ until the next scattering from the distribution in Equation \eqref{eq:exponential_distr}. We then compute the time until the first cell change event, $\Delta t_\tn{CC}$, and the first kinematic drift event, $\Delta t_\tn{KD}$, assuming that the packet continued in a straight line. We also determine the time until the next hydro-evolution event, $\Delta t_\tn{HE}$, based on $t_0$ and the length of the latest hydro-timestep, $\delta t$. The shortest of these four durations, which we call $\Delta t_\tn{min}$, determines which event will actually happen. For all events, $\Delta t_\tn{min}$ is used to first propagate the packet (via Equations \eqref{eq:r_of_t} and \eqref{eq:mu_of_t}) and then to increment $t_0$. If $\Delta t_\tn{min}=\Delta t_\tn{KD}$, only the position and angle are updated. If $\Delta t_\tn{min}=\Delta t_\tn{CC}$, then the packet switches host cell. If $\Delta t_\tn{min}=\Delta t_\tn{SC}$, then a scattering is computed. If $\Delta t_\tn{min}=\Delta t_\tn{HE}$, then the packet gets frozen after the propagation until after the next hydro-step (which gets computed when $\Delta t_\tn{min}=\Delta t_\tn{HE}$ for all photon packets). Once the appropriate action has been performed, we calculate the new local $t_\tn{mf}$ and the algorithm starts again.

To ensure that the photons accurately travel along with the plasma when the bulk speed is close to the speed of light, \verb|SPIRO| continuously interpolates the spatial positions of the cells and interfaces based on their locations at the start and end of the hydro-step.  More details are given in Appendix \ref{sec:prop_details}.

\subsection{Numerical validity} \label{sec:numerical_validity}

To properly model the system, all relevant length scales and timescales must be resolved. A necessary condition for resolving purely hydrodynamical behavior is the CFL-condition, which \verb|GAMMA| implements using the wavespeeds from the HLLC-solution.
To resolve the timescale of the photon evolution, the timestep must be smaller than the photon mean free time. This gives the condition $\delta t < \min[(2n^\prime_{e,i} \sigma_\tn{T})^{-1}]$, where $n^\prime_{e,i}$ is the comoving electron number density in cell $i$ and $\sigma_\tn{T}$ is the Thomson cross section. The corresponding length scale will be resolved if the optical depth across each cell satisfies $\Delta\bar{\tau} < 1$.

The macroscopically continuous nature of the radiation must also be resolved. The comoving energy content of a photon packet is $w\varepsilon^\prime$. The comoving energy content $U^{\prime}$ (excluding radiation energy) of a cell with $N_\tn{pl}$ plasma particles at comoving temperature $\Theta_\tn{pl}$ is
\begin{equation}
    U^{\prime} = f N_\tn{pl} k_\tn{B} \Theta_\tn{pl} m_e c^2,
\end{equation}
where $3/2 \leq f < 3$ depending on $\Theta_\tn{pl}$. Scattering events that increases $\varepsilon^{\prime}\to2\varepsilon^{\prime}$ are not uncommon. But if
\begin{equation}
    w\varepsilon^{\prime} > 3 N_\tn{pl} \Theta_\tn{pl},
\end{equation}
then such an event would extract more than the total thermal energy contained in the cell. We can avoid this problem by ensuring that 
\begin{equation}\label{eq:radiation_resolution}
    \frac{\varepsilon^{\prime}}{3\Theta_\tn{pl}} \ll \frac{N_\tn{pl}}{w},
\end{equation}
for every scattering event. The Lagrangian nature of the grid implies that that $N_\tn{pl}$ in each cell is conserved during hydrodynamical evolution. Thus, if all photon packets have an equal weight, the right-hand side of Equation \eqref{eq:radiation_resolution} scales linearly with the number of photon packets used.

The largest value that can be expected to occur in the left-hand side of Equation \eqref{eq:radiation_resolution} will also scale with the number of photon packets. But the scaling is very weak due to the exponential cutoff in the Wien distribution. A bigger issue is the fact that photons may move between regions of different plasma temperatures and bulk velocity. Furthermore, hydrodynamical cell splitting can reduce the local value of $N_{\rm pl}$. 
Therefore, to properly resolve the radiation, the number of photon packets need to be large enough such that $N_\tn{pl}/w \gg 1$ at the start of the simulation, with a large safety margins to account for cell splitting and photon mobility.

A way to decrease the required number of photon packets is to artificially raise the heat capacity of the plasma, as done in \citet{Lundman2018}. 
Multiplying the heat capacity of the plasma by $f_\tn{hc}\ge1$ is equivalent to increasing the number of particles by a factor of $f_\tn{hc}\ge1$ while keeping the mass and initial temperature of the fluid unchanged. If $f_\tn{hc}$ is small enough such that the radiation pressure still dominates the total comoving pressure, then this change will have no appreciable effect on the dynamics of the outflow, see \ref{sec:fake_heat_disc} for further discussion. 

To further reduce the risk of a cell transferring all of its internal plasma energy to the photons, \texttt{SPIRO} updates the local temperature, $\Theta_{\rm pl}$, of the cell after each scattering. Therefore, if a photon were to take a large portion of the internal plasma energy in a cell, then the local temperature would decrease, reducing the risk of a potential second scattering draining the cell within the same timestep. 

This trick is central to the efficiency of \texttt{SPIRO}. By updating the cell's temperature, and also its bulk plasma velocity $\beta_r$, we do not need to add source terms for the interaction to the hydro-solver. Furthermore, the timescales of Compton heating and Compton cooling of the plasma get resolved automatically, given that we already satisfy Equation \eqref{eq:radiation_resolution} and the CFL-condition.\footnote{According to Equation \eqref{eq:radiation_resolution}, a single scattering event is unable to cause significant temperature changes a cell. The CFL-condition in turn guarantees that energy and momentum deposited by multiple scattering events does not have time to leak into neighboring cells.}
These timescales can be multiple orders of magnitude shorter than the scattering timescale for high photon to-proton-ratios. They are also non-trivial to calculate from a cell's photon distribution when $\Theta_{\rm pl}$ is outside the classical regime.

The main downside of instant updates, as opposed to classical source terms, is that it becomes quite complicated to implement parallelized photon propagation and higher order time-stepping. The current version of \texttt{SPIRO} is a first order solver\footnote{\texttt{SPIRO} still uses \texttt{GAMMA}'s third order integrator (\texttt{SSPRK3}) for the hydrodynamics because of its stability.} and runs on a single CPU-core. The main benefit is the ability to use longer timesteps. This results in fewer calls to the Riemann solver and a massive reduction in the number of 'wasted' propagation steps, where no scattering occurs. The added cost of updating the cells is small since the computation of $\Theta_{\rm pl}$ and $\beta_r$ after a scattering is cheap compared to the scattering algorithm itself.

We have done several tests to verify the validity of the code. These include correct thermalization of the comoving photon distribution in an ultra-relativistic outflow, shock-tube tests, and correct adiabatic cooling as a function of radius in the optically regime. We have also compared the output of \texttt{SPIRO} with that against \texttt{radshock}, a radiation-hydrodynamical code that uses source terms instead of instant updates, and found excellent agreement between the two.

\section{Initial conditions for the internal collision}\label{sec:initial_conditions}

The initial ejecta profile has a setup that is very similar to the one used in \citet{AlamaaDaigne2025arXiv}. The jet is assumed to emit relativistic material with a constant observed mass flux, ${\dot M}$, while the observed energy flux, ${\dot E}$, varies smoothly across the ejecta from ${\dot E} = 10^{52}~$erg~s$^{-1}$ to ${\dot E} = 10^{53}~$erg~s$^{-1}$. The variation in ${\dot E} = 10^{52}$ leads to variations in the terminal Lorentz factor, $\Gamma_\infty \equiv {\dot E}/{\dot M} c^2$, which ranges from $\Gamma_{\infty} = 40$ to $\Gamma_{\infty} = 400$. The region where $\Gamma_{\infty} = 40$ ($\Gamma_{\infty} = 400$) will be referred to as the slow (fast) shell. At the start of the simulation, the comoving mass density, $\rho^\prime$, the comoving pressure, $p^\prime$, and the bulk Lorentz factor, $\Gamma$, in each cell are set according to the hydrodynamical fireball equations \citep[see equations (2), (3), and (5) in ][]{AlamaaDaigne2025arXiv}. We consider a fully ionized unmagnetized hydrogen plasma, consisting of an equal number of free protons and electrons. Furthermore, we assume spherical symmetry\footnote{The assumption of spherical symmetry in an ultra-relativistic jet is valid as long the jet properties do not vary strongly within an opening angle of $\sim 1/\Gamma$ to the line-of-sight.} and fix the velocity of the cell interfaces such that the mass flux across them are identically zero and the prescription becomes Lagrangian. The photons are initiated in a thermal Wien distribution at the local temperature, with isotropically distributed directions in the local comoving frame. 

In contrast to \citet{AlamaaDaigne2025arXiv}, the simulated region is twice as wide ($\Delta r = 0.3~$light-seconds) to allow the RMSs to reach smaller optical depths. Furthermore, the photon-to-proton ratio, $\zeta\equiv n_\gamma/n_p$, is for simplicity assumed constant across the grid at the start of the simulation, and we use a value that is typical for GRBs: $\zeta = 10^5$ \citep{Bromberg2011b}. Here, $n_\gamma$ and $n_p$ are the number densities of photons and protons, respectively. Lastly, the simulation begins further out at $r = 1.5 \times 10^{12}~$cm to save computational time. This starting radius corresponds to $r_{50}/r = 16.1$, where $r_{50}$ is the radius where 50\% of the Monte Carlo photons have made their last scattering as found by the simulation.

\begin{figure*}[ht!]
    \centering
    \includegraphics[width = 1.9\columnwidth]{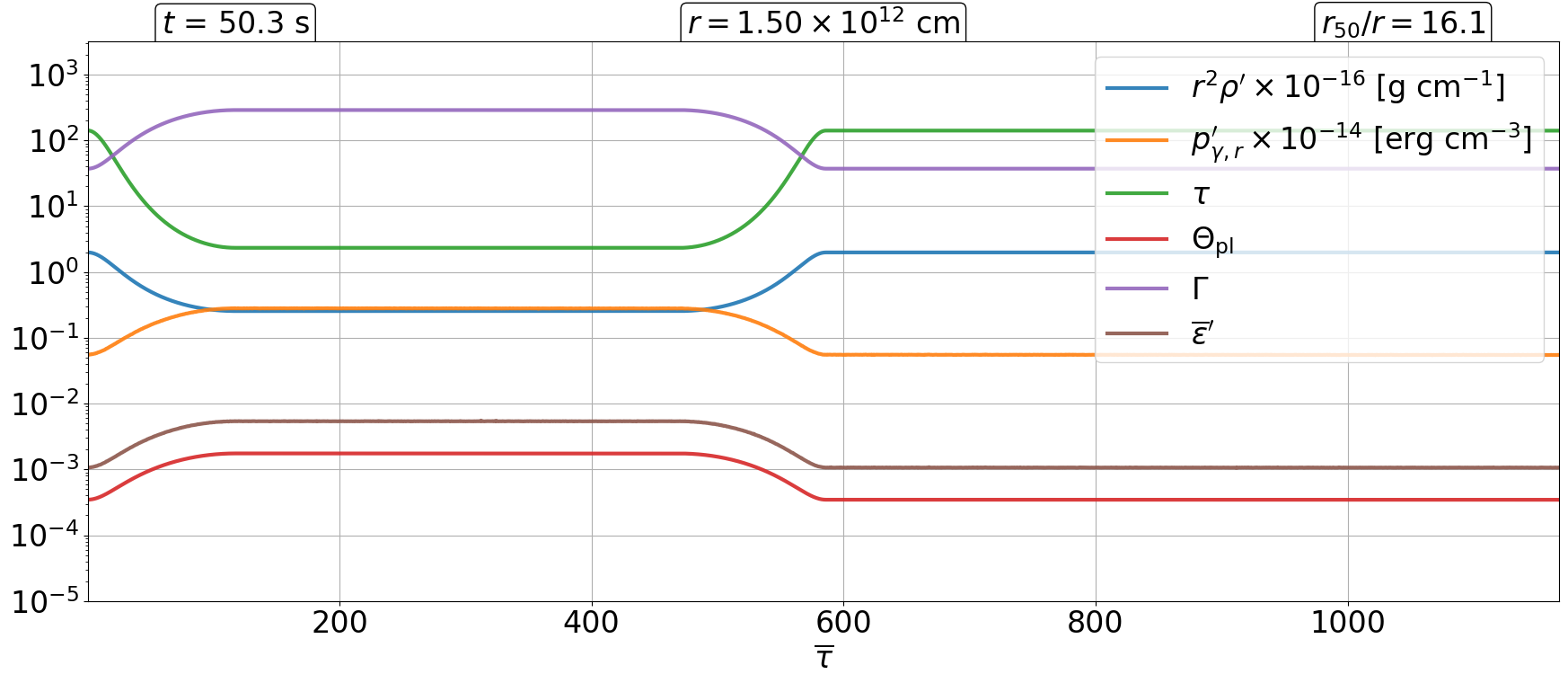}
    \caption{Initial ejecta profile across the simulated domain, with the properties plotted given by the legend. The variation in $\Gamma$ will lead to an internal collision below the photosphere with a reverse and forward RMS being formed. The smooth change in the properties at the left-hand side of the simulated domain assures continuity across the periodic boundary. The value of $r$ given is the value for the center of the grid, although the width $\Delta r \ll r$ assures negligible variation in $r$.} 
    \label{fig:init_plasma_profile}
\end{figure*}

Figure \ref{fig:init_plasma_profile} shows various ejecta properties at the start of the simulation. The properties shown are: the expansion normalized comoving mass density $r^2\rho^\prime$; the comoving photon pressure in the radial direction $p_{\gamma,r}^\prime$, 
the optical depth $\tau$, the dimensionless comoving plasma temperature $\Theta_\tn{pl}$, 
the bulk Lorentz factor $\Gamma$,  and the mean comoving photon energy $\overline{\varepsilon}^\prime$ (in units of $m_e c^2$). The optical depth is calculated in the Thomson limit as $\tau = n_e^\prime \sigma_{\rm T}r/\Gamma$ \citep{Beloborodov2011}, and we refer to regions where $\tau<1$ as being locally optically thin. Note that $\tau$ is local quantity, since it depends on $n_e^\prime$ and $\Gamma$ which can vary strongly. 
The simulation starts at $t = 50.3~$s, which is the time that has elapsed in the central engine frame since the front edge of the slow shell was emitted. The adiabatic fireball acceleration is still ongoing at this time, most noticeably in the fast shell, which means that $\Gamma$ has not yet reached its terminal value.

To better show the shock structure, all quantities are plotted in log scale and against the instantaneous Thomson opacity coordinate
\begin{equation}
    \bar{\tau}(r, t) = \int_{r_\tn{in}(t)}^r n_e(\tilde{r}, t)\sigma_\tn{T} d\tilde{r},
\end{equation}
where $r_\tn{in}(t)$ is the inner boundary of the simulation domain at time $t$ and the integral is over the static density profile evaluated at a fixed time $t$. In the optically thick regime, where the scattering time is much shorter than the dynamical time, i.e., $\tau \gg 1$, the distance that a radial photon moves through the plasma between two scattering events is of the order $\bar{\tau}\simeq 1$ \citep{RybickiLightman1979}.

The simulation domain consists of 2000 cells and each cell contains the same amount of mass, which makes the cells initially equidistant in $r$ since ${\dot M}$ is constant. We use periodic boundary conditions for both the hydrodynamics and the radiation. For the radiation, this means that photon packets that escape the simulation domain on one side, enter on the other side with their energy and weight unchanged. A periodic boundary is motivated since numerical simulations of GRB jets find rapid variability in the terminal Lorentz factor on time scales similar to what we consider here \citep[e.g.,][]{Gottlieb2019, Gottlieb2020}.

\begin{figure*}[ht!]
    \centering
    \includegraphics[width = 1.9\columnwidth]{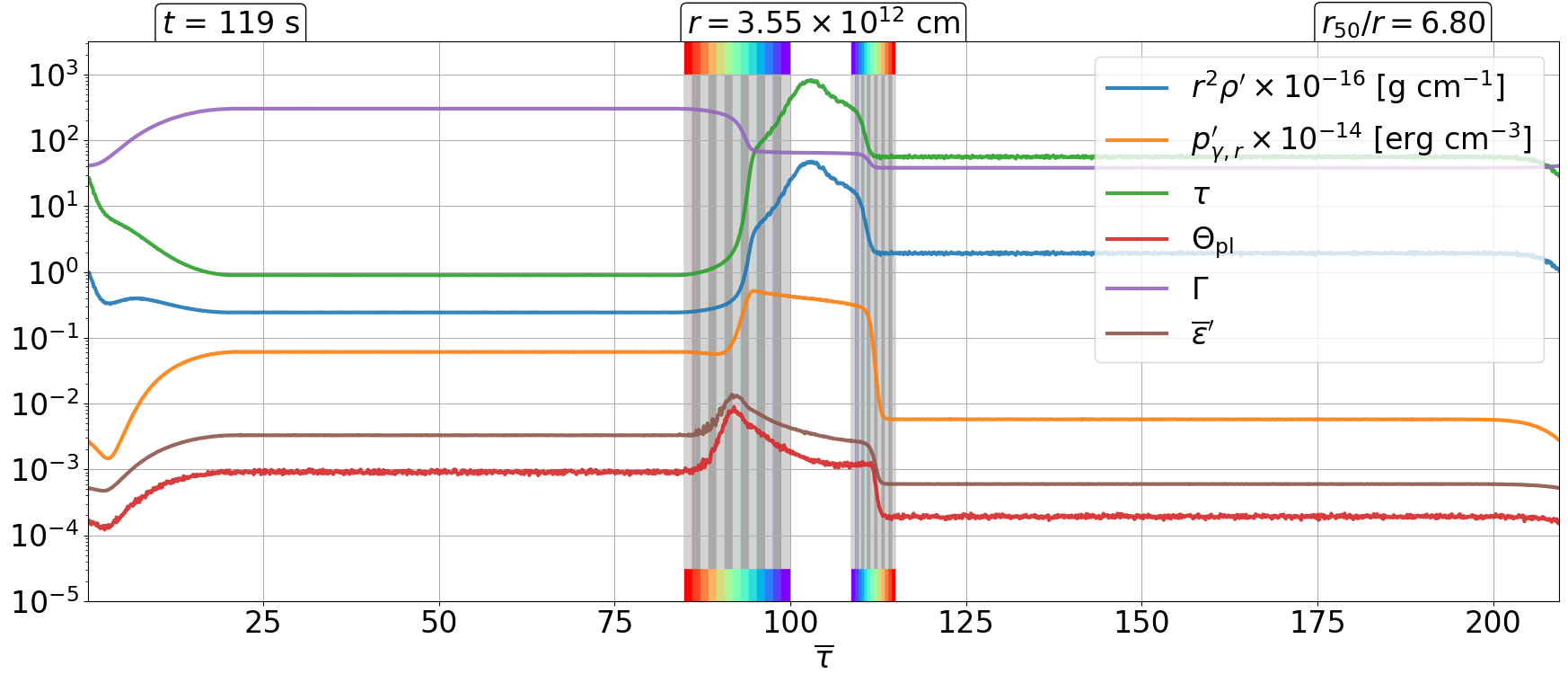}
    \includegraphics[width = 1.9\columnwidth]{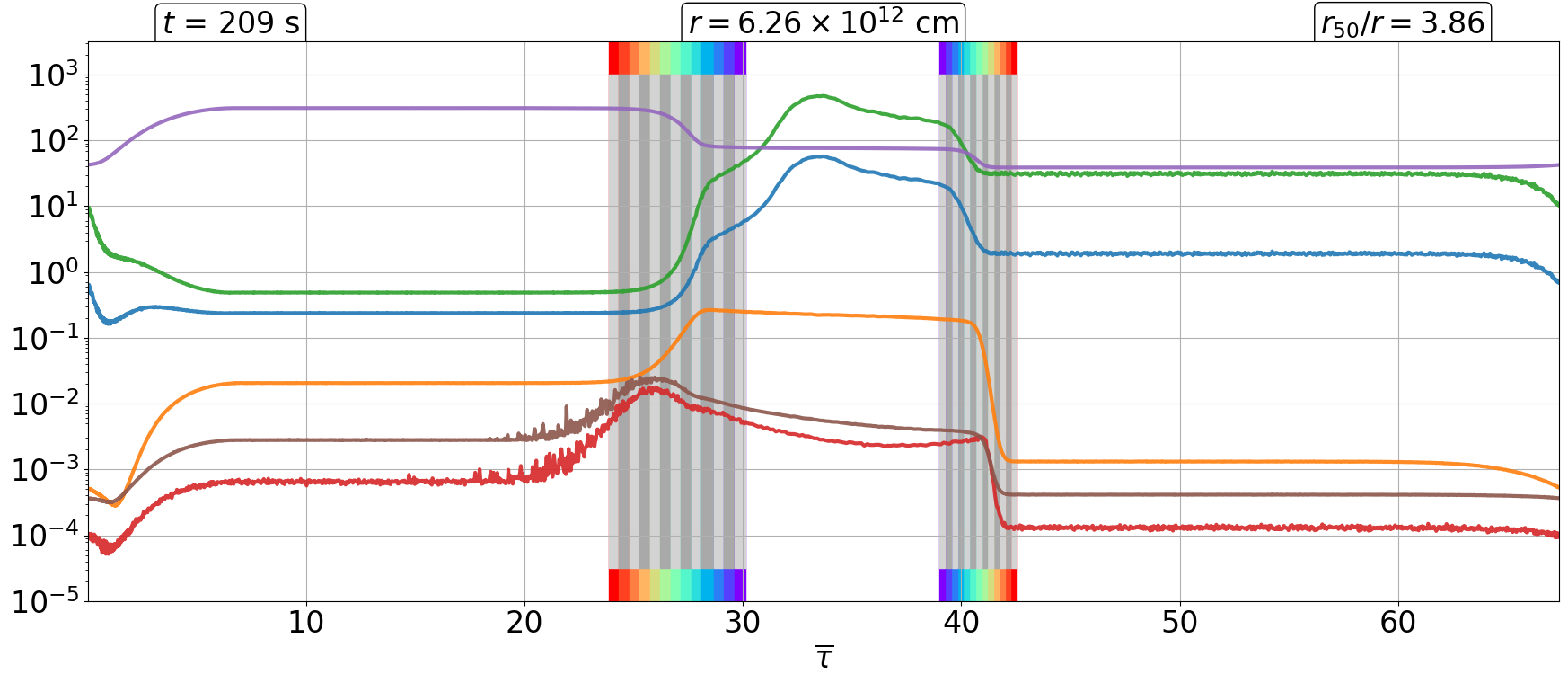}
    \includegraphics[width = 1.9\columnwidth]{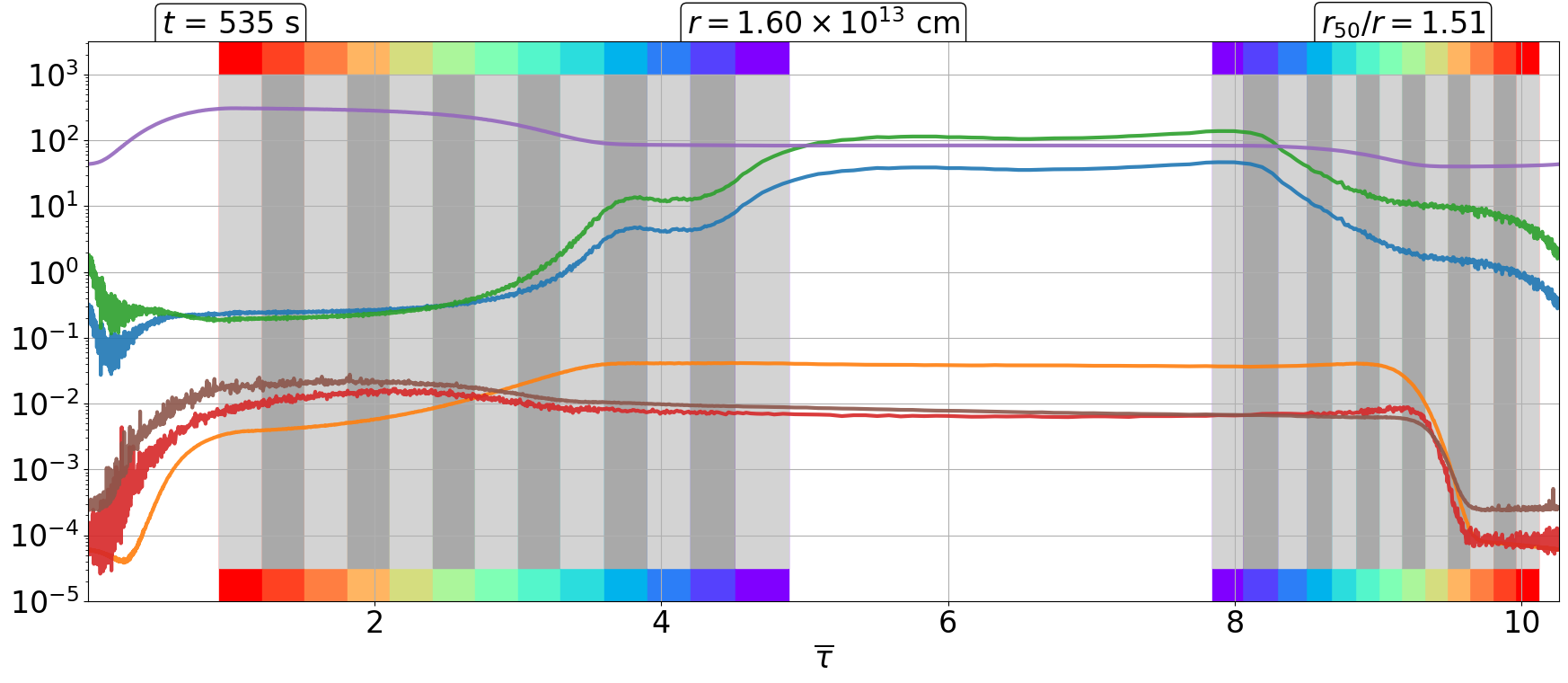}
    \caption{Ejecta profiles at $t = 119~$s (top), $t = 209~$s (middle), and $t = 535~$s (bottom) for the properties given in the legend in the top panel. At $t = 119~$s, the two RMSs have just formed and the shocks are several mean free paths wide. At $t = 209~$s, the two shocks have reached partway through their respective upstreams. High-energy photons from the reverse shock escape far upstream as a precursor. At  $t = 535~$s, both shock transition regions are very wide, covering a large portion of the simulated domain. In all snapshots, $\tau$ varies by almost three orders of magnitude across the ejecta.}
    \label{fig:plasma_profiles}
\end{figure*}

We perform two simulations in this paper, which are identical in their setup as explained thus far. They only differ in the number of Monte Carlo photon packets and the value of $f_{\rm hc}$ used. The higher (lower) resolution simulation starts with $2\times 10^5$ ($2\times 10^4$) photon packets in each cell for a total of $4\times 10^8$ ($4\times 10^7$) photon packets. To ensure that the condition in equation \eqref{eq:radiation_resolution} holds, we use an artificial heat capacity of $f_{\rm hc} = 100$ ($f_{\rm hc} = 1000$) for the higher (lower) resolution simulation, which gives $N_{\rm pl}/w = 200$ before cell splitting. The higher-resolution simulation is used to obtain the results relating to the comoving evolution presented in Figures \ref{fig:plasma_profiles} and \ref{fig:comoving_spectra}. The simulation with lower resolution is used to obtain the results relating to the observed signal presented in Figures \ref{fig:last_scattering}, \ref{fig:light_curve}, and \ref{fig:observed_spectrum}, where the lower resolution allows us to evolve the simulation far above the transparency radius, reaching $r_{50}/r \approx 0.016$. We verified that up until the end time of the high resolution simulation, the photon distributions in the two simulations were practically identical apart from numerical noise, see section \ref{sec:fake_heat_disc} for further discussion.


\section{Results}\label{sec:results}

We now present the results for the subphotospheric shell collision introduced above. Because the same radiation both mediates the shocks and forms the observed emission, the problem requires a fully time-dependent, self-consistent radiation-hydrodynamic treatment. \texttt{SPIRO} allows us to follow this evolution directly and connect the shock dynamics to the emergent signal.

\subsection{Comoving plasma evolution}

When the simulation starts, the faster shell crashes into the slower shell, resulting in an adiabatic compression in the central region that increases the density and the pressure. Once the pressure becomes comparable to the incoming ram pressure, two shocks are launched. These shocks provide a converging velocity field which allow the photons to gain energy through bulk Comptonization. The steepness of the velocity gradient, the large number of photons per proton, and the high optical depth causes the shocks to be mediated by the radiation.

In the first snapshot, at $t=119~$s, the two shocks have just formed from the collision of the two fireball shells. These shocks are several photon mean free paths wide, which makes it very unlikely for a photon to cross the entire transition region without scattering. The downstream between the shocks has been compressed, as can be seen by the elevated density in that region. Since the photon packets travel between cells, sharp transitions in the plasma are smoothed out. 
Due to the large variations in both density and Lorentz factor, the local optical depth varies by almost three orders of magnitude. While the reverse shock upstream is almost optically thin with $\tau \sim 1$, the downstream is extremely optically thick with $\tau \lesssim 10^3$. A void of rarefied plasma is starting to form at the inner edge of the fast shell, farthest to the left in the simulation region. The spatial extent of this region will grow quickly over time. Note that since we are plotting against opacity $\bar{\tau}$, this expansion only shows up as a confined dip in density.

In the second snapshot, at $t=209~$s, the reverse and forward shocks have propagated partway through their respective upstream regions. Both the plasma temperature $\Theta_\tn{pl}$ and mean photon energy $\overline{\varepsilon}^{\prime}$ have decreased slightly in the upstream regions compared to the first snapshot, due to adiabatic cooling.
The optical depth in the far upstream, $\tau_u$, has decreased to 0.5 for the reverse shock.
This allows photons to leak out as a precursor in front of the reverse shock, significantly broadening the shock transition region. This is not the case for the forward shock, where the higher $\tau_u$ allows the photons to accelerate the plasma more efficiently.

It is interesting to note that the reverse shock upstream local optical depth is below unity at this time. However, we see no indication of a subshock in the simulation, which is expected to occur at low optical depths \citep{Ito2020, LundmanBeloborodov2021}. This indicates that the radiation pressure is high enough even in these regions of low optical depth to mediate the shock completely.

In the third snapshot, at $t=535~$s, the outflow has reached $r_{50}/r=1.5$. The global decrease in opacity due to the expansion has caused both shocks to widen significantly as expected \citep{Levinson2012}. The forward shock has propagated through almost all of the slow shell. The main compression for the reverse shock still occurs well inside the fast shell, starting at ${\bar \tau} \approx 3$. However, the precursor photons have been able to propagate all the way to the void at the left edge of the simulation domain. Arguably, the shock transition region therefore extends all the way from ${\bar \tau} \lesssim 1$ to ${\bar \tau} \approx 4$, which corresponds to a radial span of $3\times10^9~$cm. Although the photons are mostly free streaming due to the low value of $\tau$ in the region ${\bar \tau} < 3$, the plasma is still tightly coupled to the photons due to the very high value of $\zeta$. This is clear from the high plasma pressure in this region, which is supplied by shock heated photons Compton scattering with the plasma.

Not even at this stage, where $\tau_u \approx 0.2$ ($\tau_u \lesssim 10$) for the reverse (forward) shock, do we see any indication of a subshock in the simulation. It is interesting to note that $\tau$ increases by a factor of $\sim 50$ ($\sim 10$) across the reverse (forward) shock transition. Therefore, even if a collisionless shock forms in a region of moderately low optical depth $\tau \sim 0.1$, Compton scattering could potentially play an important role in the formation of the emitted spectrum and the shock structure. In such a case, loads of pairs would likely be produced increasing $\tau$ even further (see section \ref{sec:disc_optically_thin_RMS}).

\def \scaledwidthR {1.07471647765}
\def \scaledwidthF {0.925283522348}
\begin{figure*}[ht!]
    \centering
    \includegraphics[width = \scaledwidthR\columnwidth]{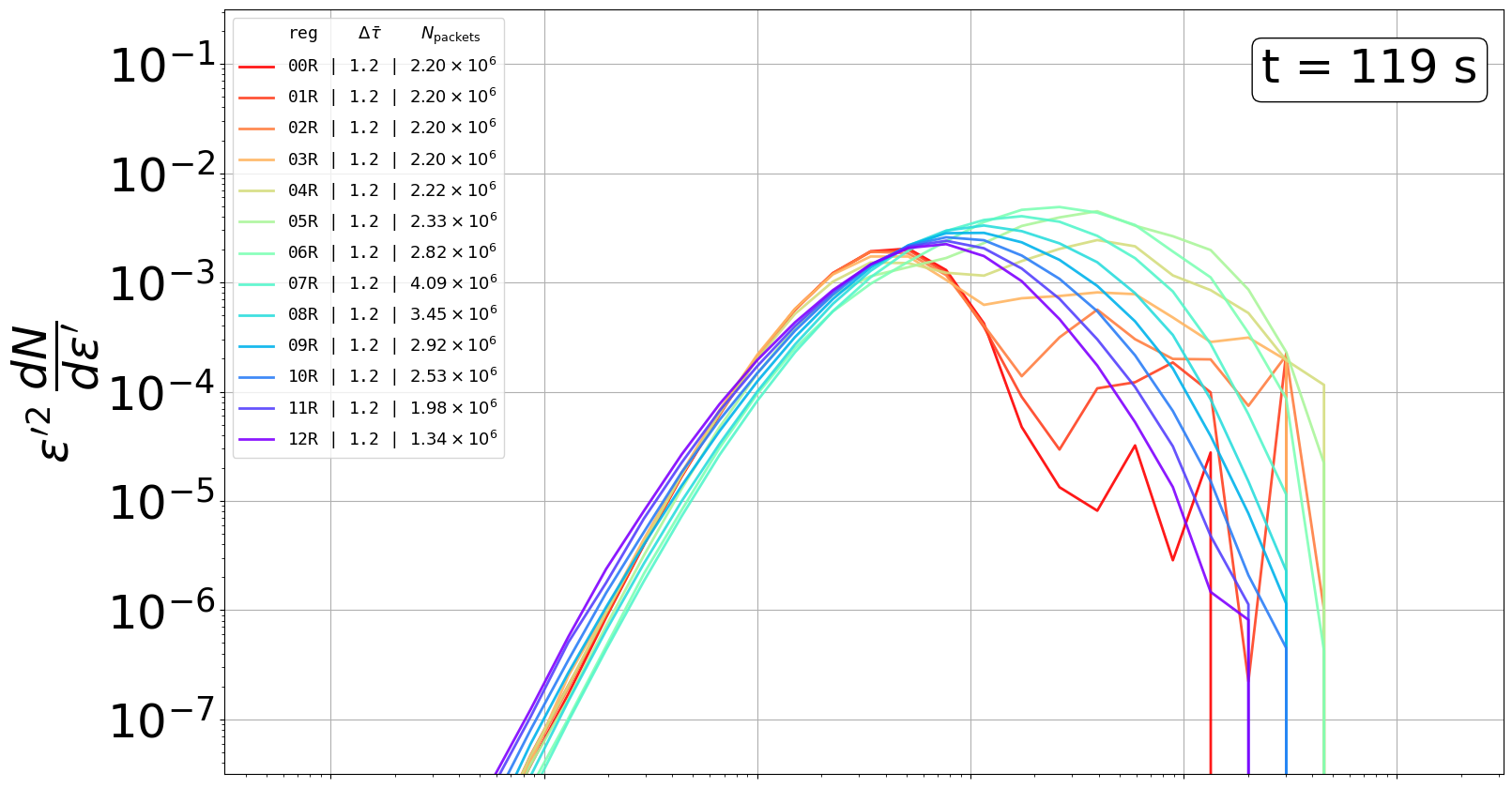}
    \includegraphics[width = \scaledwidthF\columnwidth]{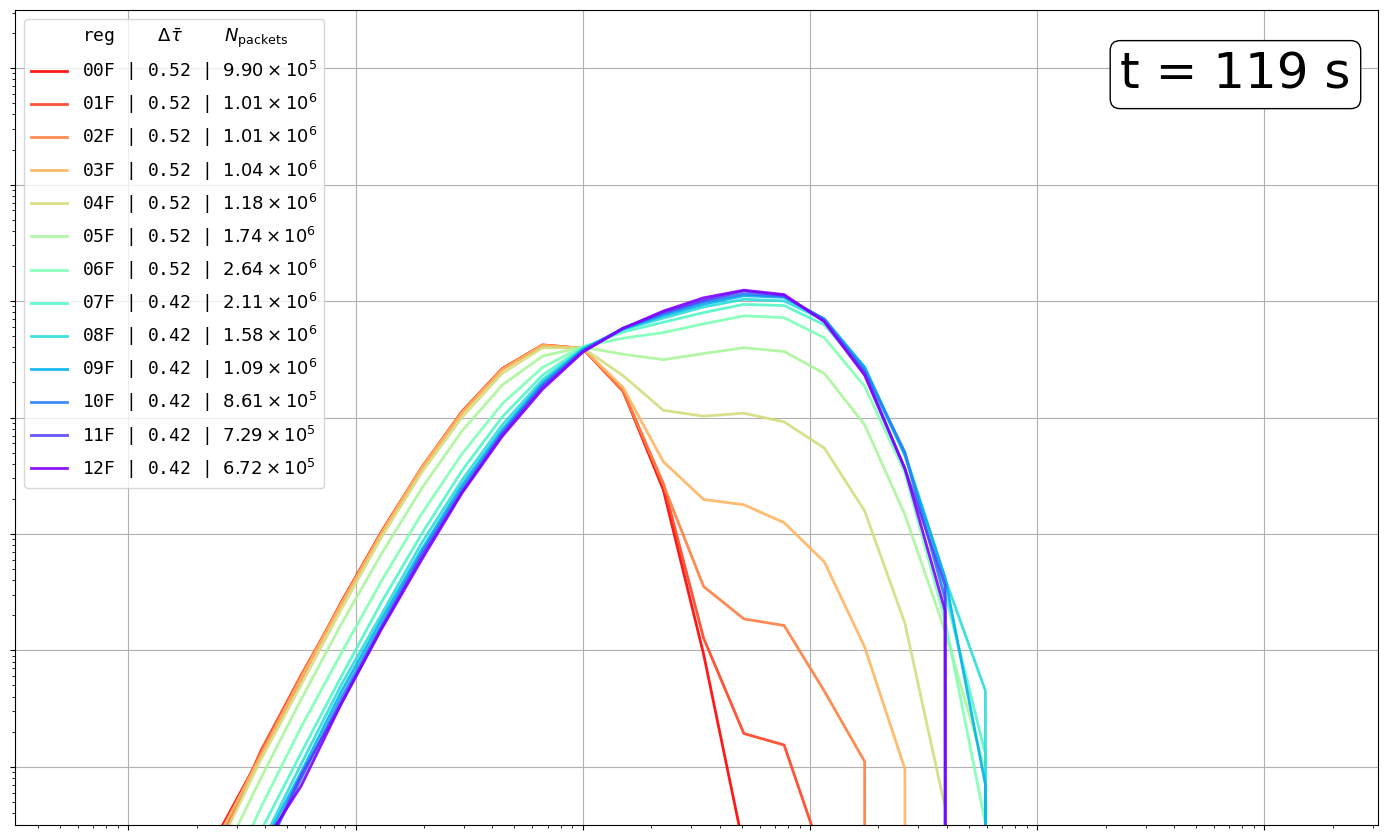}
    \includegraphics[width = \scaledwidthR\columnwidth]{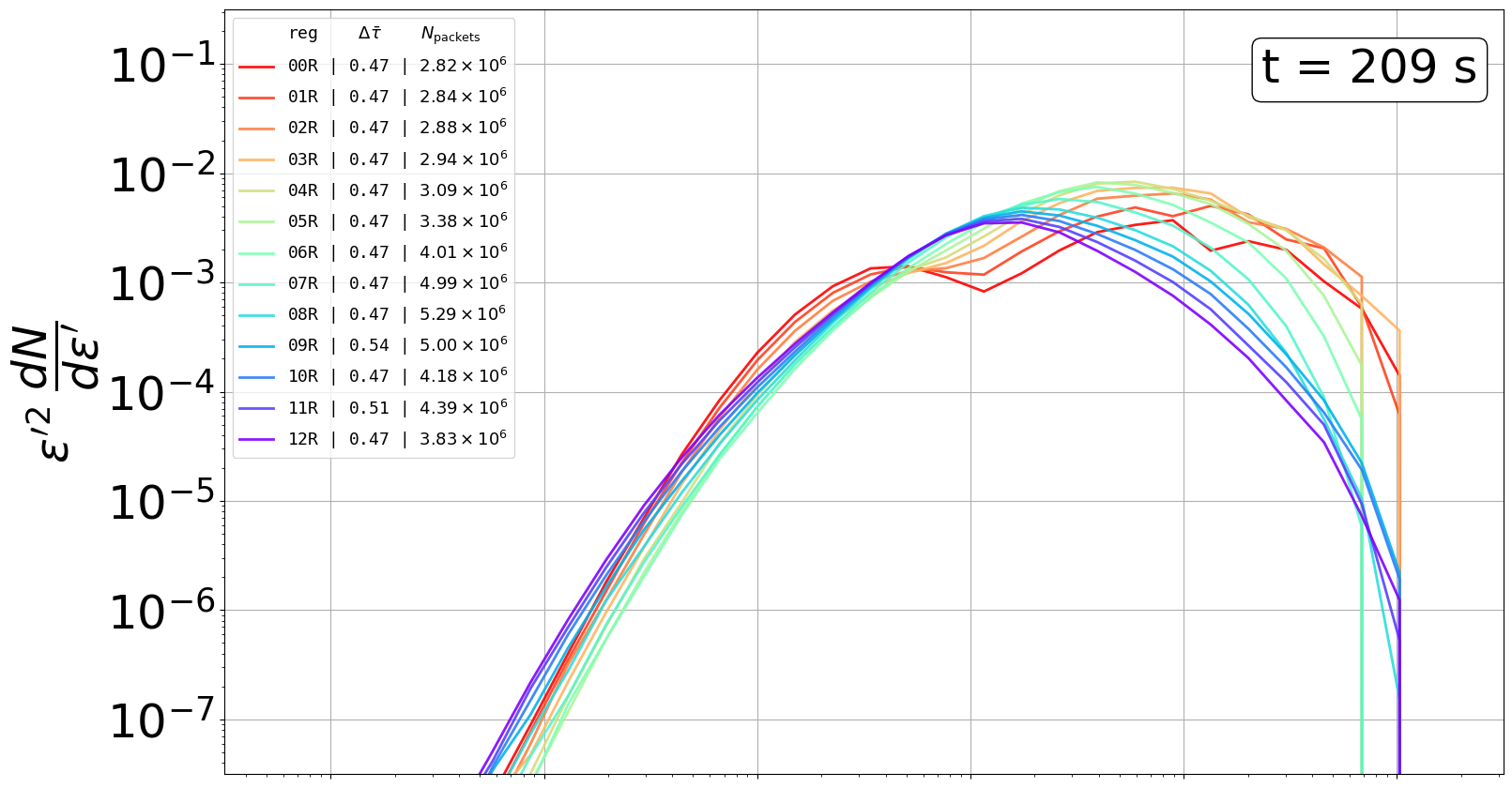}
    \includegraphics[width = \scaledwidthF\columnwidth]{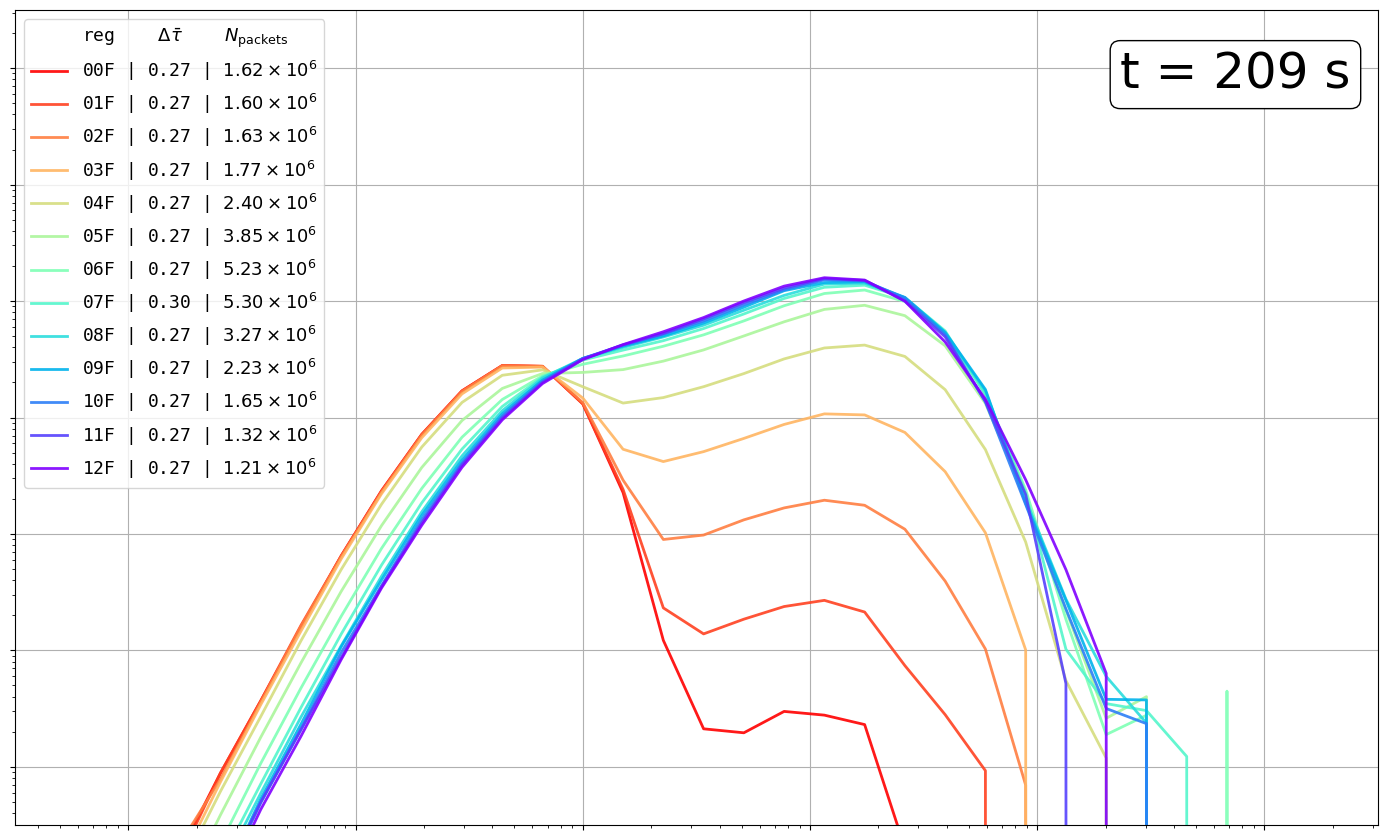}
    \includegraphics[width = \scaledwidthR\columnwidth]{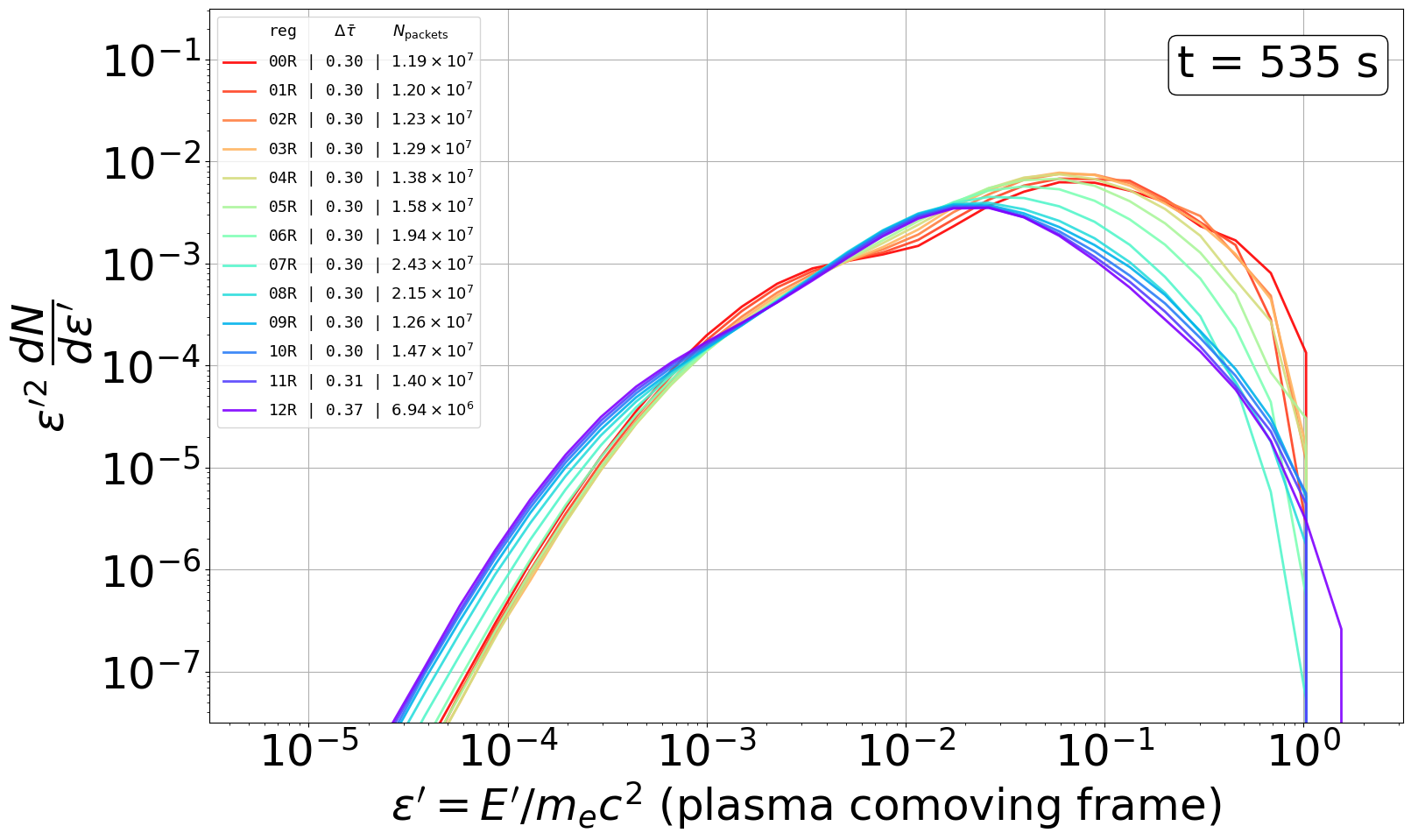}
    \includegraphics[width = \scaledwidthF\columnwidth]{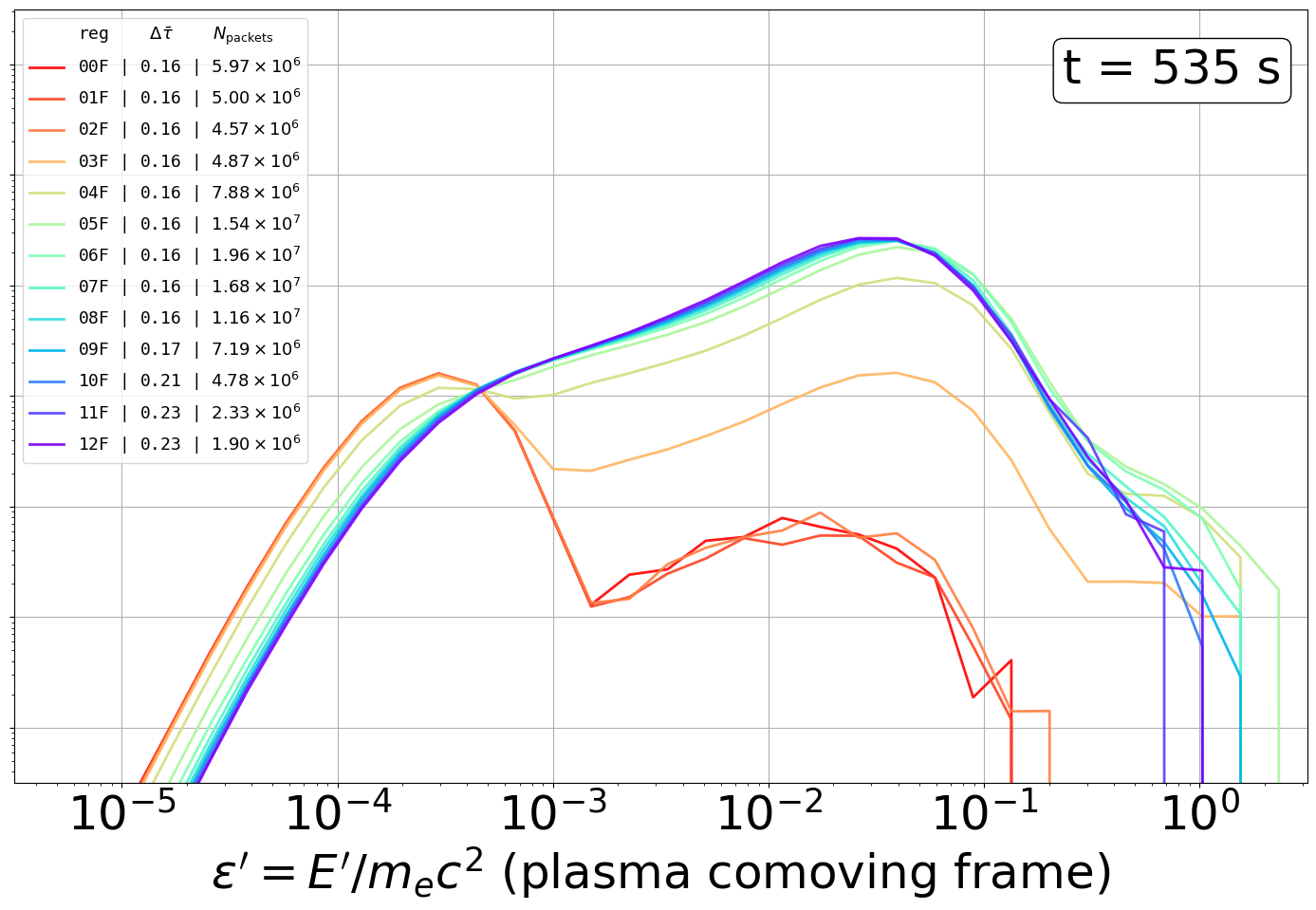}
    \caption{Individually normalized comoving spectra for the reverse shock (left) and forward shock (right) at $t = 119~$s (top), $t = 209~$s (middle), and $t = 535~$s (bottom). The spectra are evaluated in the regions marked with the corresponding colors in Figure \ref{fig:plasma_profiles}, with red evaluated farthest upstream and purple farthest downstream. The $\Delta \bar\tau$-width and number of photon packets in each region are displayed in the legend of each snapshot. The photon distributions broaden when they reach lower optical depth and are highly non-thermal at $t = 535~$s.}
    \label{fig:comoving_spectra}
\end{figure*}

\subsection{Comoving spectral evolution}
The comoving spectra inside the regions marked by the gray rectangles in Figure \ref{fig:plasma_profiles} are shown in Figure \ref{fig:comoving_spectra}. The left and right-hand panels show the photon distributions around the reverse and forward shock, respectively, ranging from red (farthest upstream) to purple (farthest downstream). Each spectrum has been normalized individually such that $\int_0^\infty (dN/d\varepsilon^\prime)d\varepsilon^\prime = 1$, where $dN/d\varepsilon^\prime$ is the photon spectral number density.


Photons are heated adiabatically together with the plasma in the initial compression. Therefore, the photon distributions around the two shocks remain quasi-thermal at early times. When the speed gradient steepens, bulk Comptonization becomes efficient and photons begin to reach higher energies. This leads to the spectrum broadening.

At $t = 119~$s, the photon distributions are still quite narrow. The forward shock red spectrum is completely thermal at this time. Thus, this region is sufficiently far upstream as to have no knowledge of the incoming shock. The reverse shock red spectrum contains a subdominant population of shock-heated photons, which indicates that the reverse shock transition region is broader already in the first snapshot. 

The forward shock upstream is much colder than the reverse shock upstream due to the lower initial internal energy in the slow shell. The velocity gradient, $d\beta_\tn{RMS}/d{\bar \tau}$, across the forward shock is rather shallow, leading to a low maximum energy $\varepsilon_{\rm max}^\prime \sim 5 \times 10^{-3}$. Here, $\beta_\tn{RMS}$ is the dimensionless velocity of the advected plasma in the shock rest frame, whose gradient over optical depth determines the maximum photon energy obtainable in the shock \citep{Samuelsson2022}.

At $t = 209~$s, the spectra have broadened. The reason for the broadening is two-fold. Firstly, the incoming photons become colder with time as the two upstreams cool adiabatically. Secondly, $\varepsilon_{\rm max}^\prime$ increases with time. The increase of the maximum energy is partly due to the shock speed increasing as it travels towards regions of lower density \citep{Sakurai1960}. It is also partly due to the the photon mean free path increasing when the density decreases, leading to an increase in $d\beta_\tn{RMS}/d{\bar \tau}$ \citep{AlamaaDaigne2025arXiv}. This is in contrast to photon poor RMSs, where $\varepsilon_{\rm max}^\prime$ decreases at low optical depth due to enhanced bremsstrahlung emission \citep{Ioka2019, Ito2020}.

The reverse shock red spectrum contains a large number of shock-heated photons. The high-energy photons can reach further into the upstream due to Klein-Nishina suppression of the cross section. This high-energy precursor can also sprinkle pairs into the upstream ahead of the shock \citep{Lundman2018}. However, the plotted photon distribution is a bit misleading. The high-energy radiation that escapes upstream from the RMS is highly anisotropic and most photons travel in the same direction. Plotting the same spectrum in the photon center-of-momentum frame instead drastically reduces the number of high-energy photons with $\varepsilon \sim 1$ available for pair production.

The reverse shock purple spectrum exhibits a high-energy power law above the peak. The power law is generated by high-energy photons downscattering on the colder electrons and extends from $\varepsilon^\prime \approx 1/N_{\rm sc}$ to $\varepsilon_{\rm max}^\prime$, where $N_{\rm sc}$ is the number of scatterings the average photon has experienced since it passed the point of maximum velocity gradient \citep{Alamaa2024_intrapulse}.

At $t = 535~$s, the spectra around both the reverse and forward shocks are highly non-thermal, spanning more than three orders of magnitude in energy. They both consist of a broad low-energy power law with a spectral hardening at low energies, and a high-energy power law with a cutoff. Although both shock regions are very wide compared to the simulated region, they are much narrower in terms of ${\bar \tau}$ compared to the earlier snapshots. Therefore, $d\beta_\tn{RMS}/d{\bar \tau}$ is very steep and photons can gain a lot of energy in each scattering.

\begin{figure}
    \centering
    \includegraphics[width = \columnwidth]{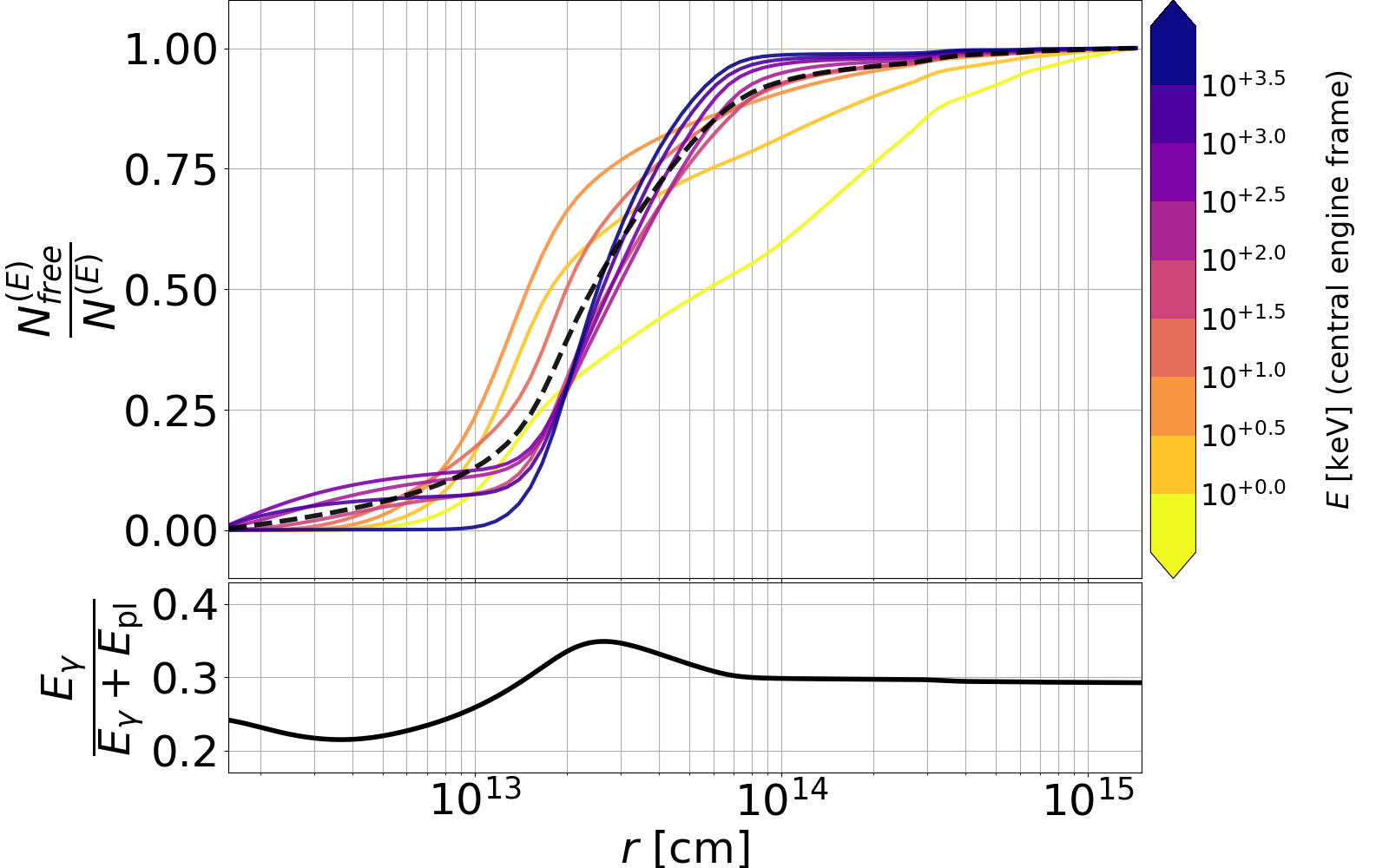}
    \caption{Cumulative distributions in different energy bands for the last scattering positions of the Monte Carlo photons. The photons that decouple very early on are situated in the reverse shock upstream, where the local optical depth is low. The majority of photons decouple once the forward shocks has reached the front edge of the simulation region.
    It is clear that photons decouple over a large range of radii.}
    \label{fig:last_scattering}
\end{figure}
\begin{figure}
    \centering
    \includegraphics[width = \columnwidth]{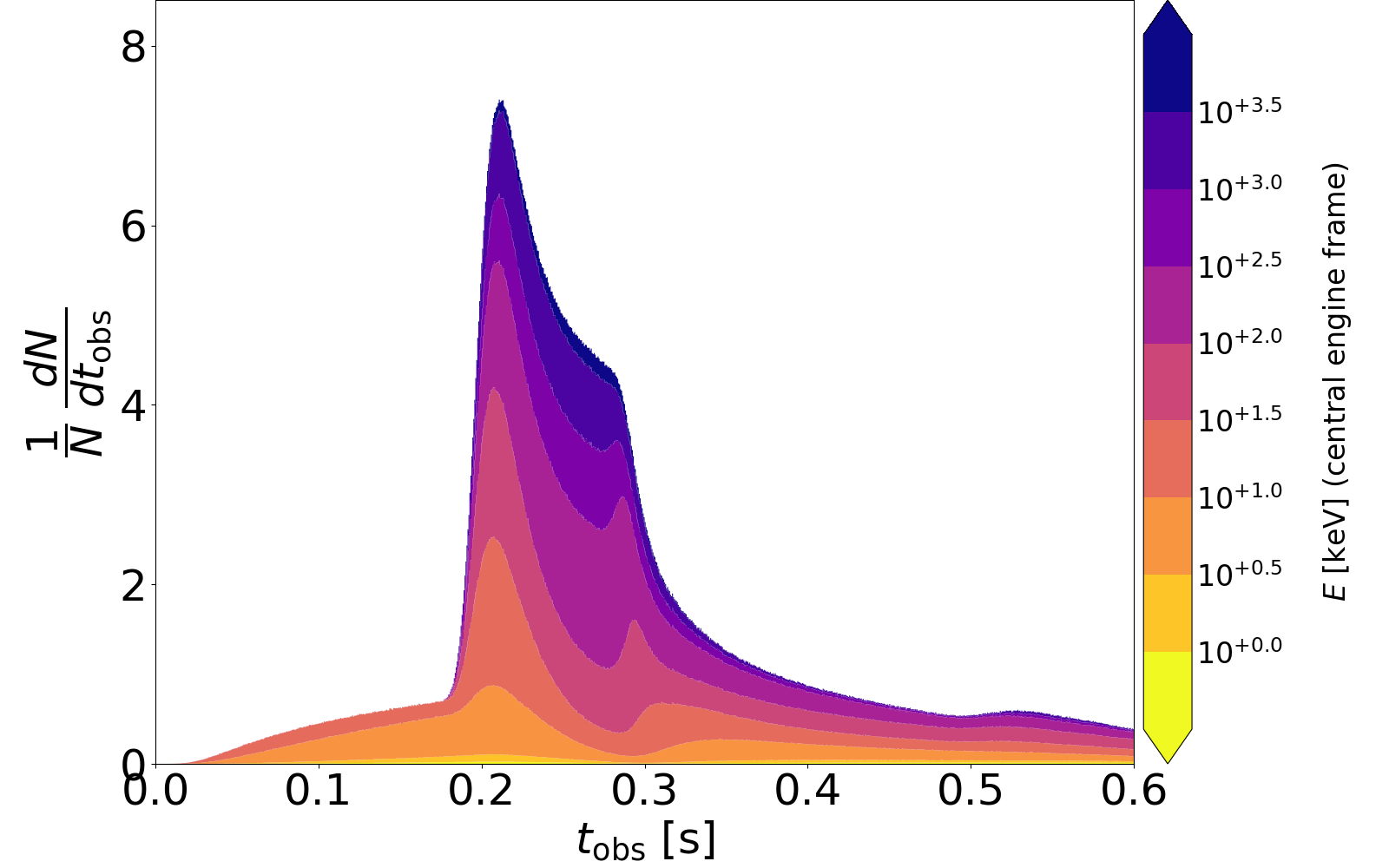}
    \caption{Central engine frame light curve, split into the same energy bands as the top panel. The first photons to reach the observer are thermal photons from the forward shock upstream. The main pulse at $\sim 0.2$~s consists of shock heated photons, while the second peak at $\sim 0.3~$s are thermal photons from the reverse shock upstream.}
    \label{fig:light_curve}
\end{figure}
\subsection{Observed signal}

Figure \ref{fig:last_scattering} shows the cumulative distributions of the last scattering positions for the Monte Carlo photon packets. The Monte Carlo photons organically decouple from the plasma at the radius and angle where they happen to make their last scattering. It is clear from the figure that photons decouple over a large range of radii, as already pointed out in several other works \citep{Peer2008, Beloborodov2011}. This motivates the use of $r_{50}$ instead of $r_{\rm ph}$, which implicitly suggests that the photosphere is a single surface.

The first photons to decouple belong not to the forward shock upstream but to the reverse shock upstream. The reverse shock upstream is only mildly optically thick at the start of the simulation and some photons become free streaming very early on. However, as can be seen by comparing with the light curve in Figure \ref{fig:light_curve}, they are not the first to arrive at the observer. Indeed, the very high optical depth in the compressed downstream region prevents all photons from crossing at early times. 

The first photons that arrive at the observer come from the forward shock upstream. These decrease in energy as a function of time due to adiabatic cooling, going from terracotta, to orange, to yellow. A rapid onslaught of photons decouple around $r \approx 1.5\times 10^{13}~$cm, when the front of the forward shock reaches the edge of the slow shell. This corresponds to the drastic increase in the light curve at $t_{\rm obs} \approx 0.2~$s. When the forward shock has reached its edge, the downstream experiences a rapid decompression with a corresponding rapid drop in the optical depth. The main pulse consists of shocked photons that decouple from the downstream when it becomes optically thin.

Interestingly, there exists a significant fraction of photons that interact with the plasma in the reverse shock upstream without ever scattering inside the reverse RMS. They experience adiabatic cooling by scattering in the upstream but decouple before the reverse shock arrives. These are the photons responsible for the second bump in the light curve at $t_{\rm obs} \approx 0.3~$s. The adiabatic cooling is clearly visible in the light curve, which shows a successive rise in all energy bands from purple to orange. This has the fascinating consequence that one might expect a multi-colored thermal post-cursor after the main peak.

\begin{figure}
    \centering
    \includegraphics[width = \columnwidth]{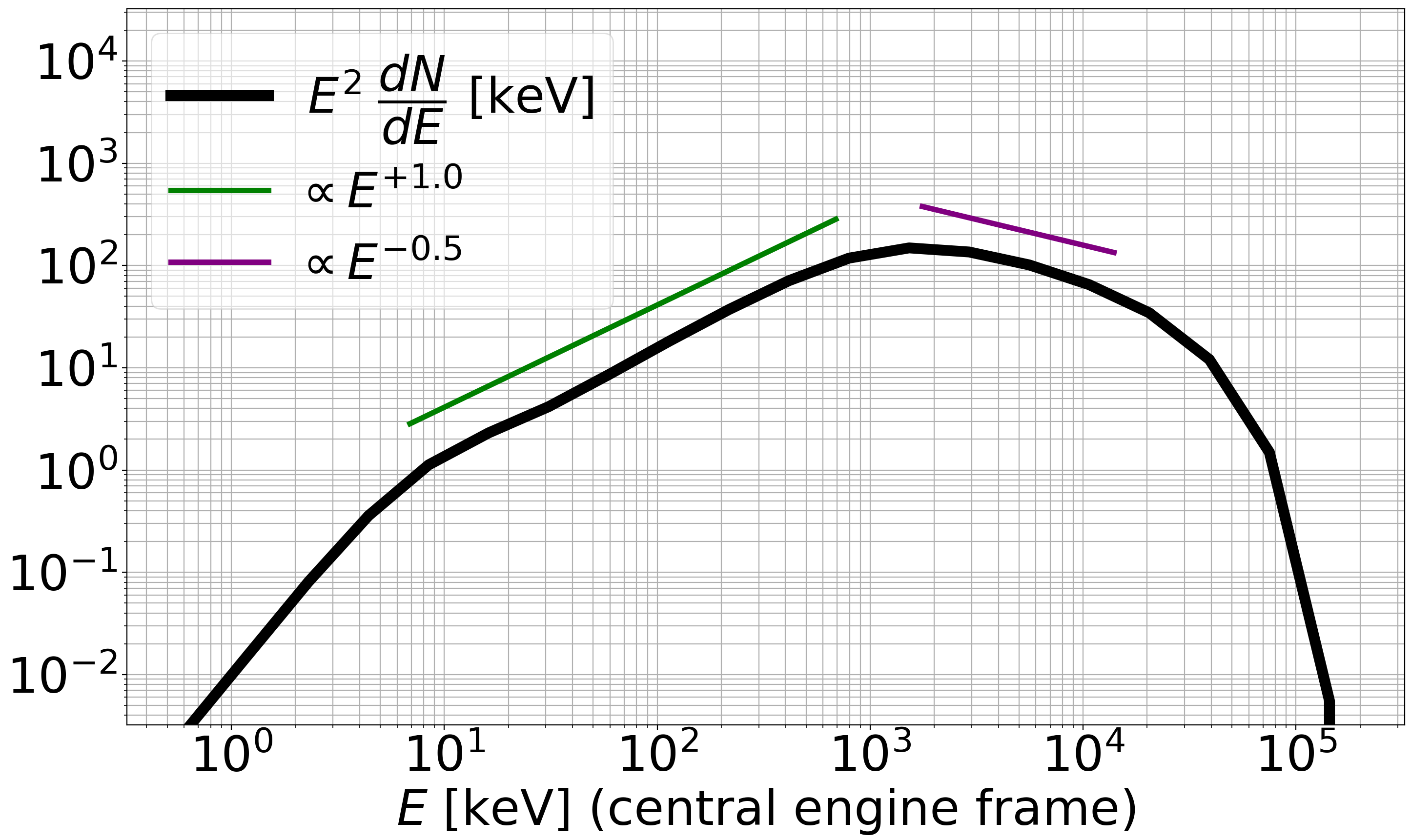}
    \caption{Normalized time integrated $\nu F_\nu$-spectrum in the central engine frame. The green and purple line show slopes corresponding to photon indices $-1$ and $-2.5$, respectively, to guide the eye. The general characteristics are a hard power law at $E \lesssim 10~$keV, a low-energy power law extending $\lesssim 2~$orders of magnitude with index $\alpha \approx -1$, a peak energy of $E_p \approx 1.5~$MeV, a high-energy power law with index $\beta \approx -2.5$, and a cutoff at $\sim 20~$MeV.}
    \label{fig:observed_spectrum}
\end{figure}

The boundary condition used in the simulation is periodic. Hence, the light should also be periodic with a period of $0.3~$s (the light-crossing time of the simulated region). Including emission from the collision in front of and behind the simulated regime changes the shape of the light curve. However, the main peak at $0.2$--$0.3~$s is largely unaffected by this (the result is to add the emission between $0.5$--$0.6~$s to the main pulse).

The lower part of Figure \ref{fig:last_scattering} shows the fraction of the total energy that is carried by radiation as a function of radius. The total energy is the sum of the energy in the plasma (internal plus kinetic), $E_{\rm pl}$, and the radiation, $E_{\gamma}$. At the start of the simulation, radiation carries $\sim 25$\% of the total energy. This fraction decreases initially as the internal energy is converted into kinetic energy of the fireball. However, the decrease is halted around $2 \times 10^{12}~$cm once the shocks form. At the peak, radiation energy constitute $\sim 35$\% of the total energy, which happens around $r_{50}$. After this point, the fraction decreases again as the compressed downstream expands and cools. As the photons decouple, the fraction asymptotes, with roughly $30$\% of the total burst energy being carried by radiation at the end of the simulation.

The time integrated $\nu F_\nu$-spectrum ($\propto E^2 \frac{dN}{dE})$ is shown in Figure \ref{fig:observed_spectrum}. The spectrum peaks at $E_p \approx 1.5~$MeV in the central engine frame and the observed spectrum would be a factor of $(1+z)$ lower due to cosmological redshift. Its shape is roughly that of a broken power law, with a low-energy photon index $\alpha \sim -1$ and a high-energy index $\beta \sim -2.5$,\footnote{The photon indices $\alpha$ and $\beta$ are defined such that $\frac{dN}{dE} \propto E^{\alpha,\beta}$.} and a high-energy cutoff at $\sim 20~$MeV. The values of the peak energy and the power-law indices are similar to those observed in GRBs \citep{Yu2016}. However, the detailed spectral shape is more complex than this overarching picture. Below $\sim 10~$keV, the spectrum hardens. This low-energy cutoff corresponds to the thermal photons decoupling from the upstream. 
At $\sim 30~$keV, the quasi-thermal component and the shock-heated component intersect, which creates a small but noticeable dip in the spectrum. Above this dip, a clean power law extends for one order of magnitude with a slope $\alpha \sim -0.9$. Above the peak, the spectral shape resembles a power law, but in reality the slope of the spectrum is changing continuously. It is better described by a smooth curvature that becomes progressively softer until it disappears completely at $\sim 100~$MeV.

\section{Discussion}\label{sec:discussion}

\subsection{Radiation-mediated shocks in the optically thin regime}\label{sec:disc_optically_thin_RMS}
One interesting result is that we find the reverse shock to be radiation-mediated with no visible subshock until very low values of $\tau_u \sim 0.1$. 
When $\tau$ is small, roughly a number fraction $\tau$ of the available photons will scatter within the next dynamical time. A photon-to-lepton fraction of $10^5$ and an optical depth of $\tau = 0.1$ then implies that each electron scatters $\sim 10^4$ times during the next dynamical time. In addition, both forward and reverse shocks show a large (order of magnitude) increase in $\tau$ over the shock transition due to the compression of the plasma, a compression which is always expected to be high in internal collisions with high radiative efficiency. 

Based on the arguments above, a natural question to ask is what would happen if the internal collision had begun in the optically thin region. The plasma might then initially be devoid of photons, or it may be embedded in a free streaming photon field emitted from regions of the jet closer to the central engine. Regardless, the collision will heat the plasma and synchrotron emission will enrich the plasma with photons. If the collision occurs at $\tau > 0.1$, the downstream will become optically thick due to compression and the radiation will be trapped. Furthermore, inverse Compton scattering would populate the plasma with high-energy photons that could increase the opacity via pair production, and the enhanced opacity can trigger violent pair production in a runaway process \citep{Beloborodov2002, Salafia2026_arXiv}. 

There are several ways this system could evolve. The collisionless shock might generate such heavy pair loading that the shock transitions to an RMS. However, once the shock becomes mediated by radiation, particle acceleration becomes inefficient and the pair production could cease. Thus, the shock might oscillate back and forth between collisionless and RMS or it may reach a quasi steady state with a collisionless subshock embedded in a wider RMS. In the least extreme scenario, the shock stays collisionless with no dynamical influence from the radiation. But even in this case, an optically thick downstream should leave an imprint on the emitted signal.

\subsection{Assumptions and model validity}\label{sec:assumptions}

In this paper, the radiation transfer is treated carefully while many other complications have been omitted. This is by design.
Keeping the model minimal but self-consistent allows us to isolate and study the behavior stemming from the complex radiation-hydrodynamical interactions.
Relying on very few initial ingredients, we obtain an emitted spectrum from first principles that is very similar to the prototypical GRB spectrum. This gives an indication that our assumptions are justified. Additional model complexity can be added iteratively once/if it is needed.

The key assumptions are a simple geometry, negligible magnetic fields, negligible neutron component, no photon or pair production, and fully ionized hydrogen plasma behaving as a single collisional fluid. In the rest of this section, we go through and discuss these assumptions in turn. We also discuss the use of artificial heat capacity.

\subsubsection{Geometry}
We considered an internal collision in 1D spherical symmetry with periodic boundary conditions. The boundary conditions used are motivated by numerical simulations of GRB jets that find the ejecta to be highly variable \citep{Lopez-Camara2014, Ito2015, Gottlieb2019}. It implies that the collision we study is just one among many others in the outflow. It also implies that the pulse presented in the Figure \ref{fig:light_curve} would constitute only a part of the GRB light curve, which can be much more variable as a whole. 

The assumption of 1D spherical symmetry can be justified as follows. In ultra-relativistic jets, the observer only sees radiation within a viewing angle $\sim 1/\Gamma$ to the line-of-sight. Thus, the jet must only be uniform over a small opening angle for the ejecta to be effectively spherically symmetric. Mixing and turbulence are features that can only be captured in higher dimensions. However, at the distances considered here, the outflow is well described by ballistic motion, which motivates the 1D approach \citep{Gottlieb2019}.

\subsubsection{Magnetic fields}
Magnetic fields likely play a role in the launching of the jet. However, whether they survive to large radii is still an open question. The role that magnetic fields play on the dynamics and emission depends on the value of the magnetization parameter, $\sigma = B^{\prime 2}/4\pi \rho^\prime c^2$, where $B^\prime$ is the comoving magnetic field strength. If the magnetization remains large, $\sigma \gtrsim 1$, then magnetic fields play a dynamically important role. Even at lower magnetization, $\sigma \sim 0.01$--$0.1$, magnetic fields can affect the radiation by forming a strong collisionless subshock \citep{Beloborodov2017}. The subshock changes the photon distribution by rapidly heating the plasma,
resulting in inverse Compton scattering and synchrotron emission behind the subshock \citep{LundmanBeloborodov2019}.

\subsubsection{Neutrons}
In this work, we assumed a low neutron-to-proton ratio, $n_n /n_p \ll 1$. When neutrons are present, inelastic scatterings between neutrons and protons can generate pions, which form a pair cascade when they decay \citep{Derishev1999}. For pion production to be possible, a large relative drift velocity between the neutron and proton components in the jet is necessary. In our presented case, the neutrons would stay coupled to the protons during acceleration \citep{Beloborodov2010}. However, a relative drift velocity can occur over the shock transition region as the neutrons are unaffected by the charge separation that decelerates the protons in the shock \citep{Beloborodov2010, Beloborodov2017}. How big of an effect this could have on the emitted spectrum depends on the neutron-to-proton ratio, the neutron opacity where the shocks occur, and the final relative drift velocity. 

\subsubsection{Photon and pair production}
Non-magnetized RMSs that occur in GRBs are photon rich, which means that photon-to-proton ratio is so high that photon production in the shock transition region can be ignored \citep{Bromberg2011b}. For $\zeta \sim 10^5$, this remains true even at low optical depth were radiative losses become important \citep{Ioka2019}. Furthermore, the two simulations presented in this paper produce no pairs according to our calculation detailed in Appendix \ref{app:pair_check}. This is line with the calculation presented in Appendix B in \citet{Samuelsson2022}, which estimates that a higher ratio of $\Gamma_\infty$ is needed for pair production to become important.
Thus, we conclude that the results presented in Section \ref{sec:results} are unaffected by the assumptions of no photon and pair production.

\subsubsection{Composition}
Considering an ionized hydrogen plasma is valid since the intense radiation field leads to all heavier elements being disintegrated at the base of the jet \citep{Murase2008, Horiuchi2012}. Furthermore, the coupling between electrons (Compton scattering targets) and protons (carriers of the bulk kinetic energy) is expected to be strong in the absence of magnetic fields and pairs \citep{Lundman2018, Levinson2020}, which justifies the single fluid approximation for the plasma.

\subsubsection{The effect of artificial heat capacity}\label{sec:fake_heat_disc}
The two simulations performed in this paper only differ in the number of Monte Carlo photon packets and the value for the artificial heat capacity used. This allows us to estimate the influence of the artificial heat capacity by comparing the output of the two simulations. Neglecting noise, which is more prevalent in the low-resolution simulation, we find that the two simulations are essentially identical in most aspects. Specifically, the comoving spectra across both RMSs are almost identical. This implies that using artificial heat capacity does not influence the observed signal in any appreciable way.

There is a difference in the plasma pressure at late times. Due to the high photon-to-proton ratio, the radiation decouples from the plasma long before the plasma decouples from the radiation. This means that the plasma temperature keeps dropping due to adiabatic cooling even in the optically thin region. When $f_{\rm hc}$ is high however, more energy transfer between photons and plasma is needed to regulate the temperature, causing the plasma temperature to freeze out too early at a value that is too high. Thus, using artificial heat capacity means that the plasma temperature in the bottom panel in Figure \ref{fig:plasma_profiles} is overestimated. Again, we stress that this has no visible effect on the observed radiation.


\section{Conclusion}\label{sec:conclusion}

In this paper, we have presented the first self-consistent radiation-hydrodynamic simulation of a subphotospheric internal collision in a GRB outflow, from the formation and propagation of forward and reverse radiation-mediated shocks all the way to photon decoupling and free-streaming towards the observer. 
The challenge is that the plasma and the radiation are strongly coupled through Compton scattering and co-evolve in a nontrivial way, and that the same photons both mediate the shocks and form the observed emission. 
Capturing this coupling requires radiation feedback on the plasma. To solve the above problem, we have developed the simulation code \texttt{SPIRO}. By including this feedback, \texttt{SPIRO} makes it possible to study the formation of the GRB emission, where steady-state approaches and post-processed radiative transfer are not sufficient.

The setup for the simulation was shown in Figure \ref{fig:init_plasma_profile} and our main results can be summarized as follows.

\begin{enumerate}
    \item The ejecta evolution as a function of time was shown in Figure \ref{fig:plasma_profiles}. Due to differences in the comoving density and the bulk Lorentz factor, we found that the reverse shock upstream becomes optically-thin at much earlier times than the forward shock upstream. Furthermore, the compression behind the shocks meant that the optical depth $\tau \equiv n^{\prime}_e \sigma_{\rm T} r/\Gamma$ varied by several orders of magnitude over the simulated domain. This led to photons making their last scattering over a vast region, stretching more than two decades in radius, as shown in Figure \ref{fig:last_scattering}.

    \item We find that the reverse shock stayed radiation mediated with no visible collisionless subshock even when $\tau_u \ll 1$, where $\tau_u$ is the optical depth in the far upstream region. 
    We argued that this was due to 1) the shock compression, which led to an increase in $\tau$ of a factor $\sim 50$ across the reverse RMS, meaning that a large portion of the shock transition region was optically thick even though the upstream was optically thin and 2) the high photon-to-lepton ratio, which kept the plasma tightly coupled to the radiation long after the majority of the photons had become free streaming. This indicates that Compton scattering should play an important role in the formation of the emitted spectrum and the shock structure even for collisionless shocks that form at $\tau \sim 0.1$. 

    \item The evolution of the comoving photon distributions around the two shocks was shown in Figure \ref{fig:comoving_spectra}. Before bulk Comptonization had become efficient, the distributions were quite narrow. However, as the shocks propagated towards lower optical depths, the photon distributions broadened significantly and developed a clear broken power law below the peak energy and a power law with a cutoff above the peak energy.

    \item The light curve was shown in Figure \ref{fig:light_curve}. The radiative efficiency was high, with $\sim 30$\% of the total burst energy being radiated. Interestingly, the light curve showed a re-brightening after the main peak due to photons having decoupled early on in the reverse shock upstream. This manifested itself as a quasi-thermal postcursor after the main emission, whose temperature decreased with time due to adiabatic cooling in the comoving upstream.

    \item The time-integrated emitted spectrum in the central engine frame was shown in Figure \ref{fig:observed_spectrum}, and consisted of a low-energy power law with index $\alpha \sim -1$, a spectral hardening below $\sim 10~$keV, a peak energy of $E_p \sim 1.5~$MeV, and a high-energy power law with index $\beta \sim -2.5$.
\end{enumerate}

\noindent Results 2--5 presented above were obtained thanks to the unique capabilities of \texttt{SPIRO}.

Within this deliberately minimal but physically motivated setup, we find that a simple internal collision below the photosphere can reproduce several key characteristics of the prompt GRB emission. These results show that time-dependent radiation-mediated shocks below the photosphere are not only physically viable, but are capable of directly linking the internal dynamics of the outflow to the observed spectral and temporal properties.

\begin{acknowledgements}
We thank Andrei Beloborodov for fruitful discussions. FA is supported by the Swedish Research Council (Vetenskapsr\aa det, 2022–00347). FR acknowledges support from the Swedish National Space Agency (2022–00205 and 2024-00128).
\end{acknowledgements}


\newpage
\appendix

\section{Klein-Nishina corrections} \label{sec:kn_corrections}
The total Klein-Nishina cross section is given by
\begin{equation}
    \sigma_\tn{KN}(\varepsilon^{\prime\prime}) = \frac{3}{4}\sigma_\tn{T}{\Biggl [}{\frac {1+\varepsilon^{\prime\prime} }{(\varepsilon^{\prime\prime})^3}}{\Biggl (}{\frac {2\varepsilon^{\prime\prime} (1+\varepsilon^{\prime\prime} )}{1+2\varepsilon^{\prime\prime} }}-\ln {(1+2\varepsilon^{\prime\prime} )}{\Biggr )}+{\frac {\ln {(1+2\varepsilon^{\prime\prime} )}}{2\varepsilon^{\prime\prime} }}-{\frac {1+3\varepsilon^{\prime\prime} }{(1+2\varepsilon^{\prime\prime} )^2}}{\Biggr ]},
\end{equation}
where $\varepsilon^{\prime\prime}$ is the dimensionless photon energy in the electron rest frame. It is a decreasing function of $\varepsilon^{\prime\prime}$ and reduces to $\sigma_\tn{T}$ in the low-energy limit, $\varepsilon^{\prime\prime}\to0$.

Consider a photon propagating through a thermal electron gas that is hot enough to have an appreciable dimensionless temperature, $\Theta = k_\tn{B} T / m_e c^2$. The Lorentz factor of an electron relative to the comoving frame of the gas, $\gamma_e$, will follow the Maxwell-Jüttner distribution
\begin{equation}
    f_\tn{MJ}(\gamma_e,\Theta)\,d\gamma_e = \frac{\gamma_e^2 \sqrt{1-\gamma_e^{-2}}}{\Theta\, \tn{K}_2(\Theta^{-1})}  \exp\lrp{-\gamma_e/\Theta}\,d\gamma_e,
\end{equation}
where $\tn{K}_2$ is the modified Bessel function of the second kind, of order 2. The corresponding thermal electron speed is given by $\beta_e = \sqrt{1-\gamma_e^{-2}}$. The photon energy in the electron rest frame depends on the the comoving photon energy, $\varepsilon^{\prime}$, via
\begin{equation}
    \varepsilon^{\prime\prime} = (1 - \beta_e\mu_e^{\prime})\gamma_e\varepsilon^{\prime},
\end{equation}
where $\mu_e^{\prime}$ is the cosine of the comoving angle between the momenta of the two particles. By averaging over all directions, we obtain the effective scattering cross section
\begin{equation}\label{eq:effective_sigma}
    \tilde\sigma(\varepsilon^{\prime}, \Theta) = 
    \int_1^\infty f_\tn{MJ}(\gamma_e,\Theta)\lrp{\frac{1}{2}\int_{-1}^1(1-\beta_e\mu_e^{\prime})\sigma_\tn{KN}(\varepsilon^{\prime\prime})\, d\mu_e^{\prime}} d\gamma_e,
\end{equation}
The inner integral can be expressed analytically using the dilogarithm $\tn{Li}_2$. During initialization, \verb|SPIRO| precomputes values of $\tilde\sigma$ in a rectangular lattice of points in $(\varepsilon^{\prime}, \Theta)$-space.\footnote{The simulations in this paper used a lattice made from 70 values of $\varepsilon^{\prime}$ and 40 values of $\Theta$. Specifically $\varepsilon^{\prime} = [0,\, 10^{-3},\dots,\,10^4]$ and $\Theta = [0,\, 10^{-2},\dots,\,10^2]$, where the values implied by the $\dots$ are log-spaced.
Note that $\tilde\sigma(\varepsilon^\prime,0)=\sigma_\tn{KN}(\varepsilon^\prime)$ and $\tilde\sigma(0,\Theta)=\sigma_\tn{T}$ for all $\Theta\ge0$.
} Those values are then used for bilinear interpolation at runtime.

To compute scattering events we need to obtain individual electron velocities. There are two sources of bias that affect the distribution of velocities among the electrons that the photon scatters with. The first is \textit{encounter rate bias}. The differential number of electrons that a photon encounters per unit time is proportional to $1-\beta_e\mu_e^{\prime}$. The photon can never interact with electrons with $\beta_e\mu_e^{\prime}=1$ since the Galilean relative velocity is zero in that case. The second source of bias comes from the energy dependence in the cross section. The photon is more likely to scatter on an encountered electron if its energy is low in the rest frame of that electron. These two effects are partially antagonistic. A high encounter rate $\beta_e\mu_e^{\prime}\approx-1$ implies a high Doppler factor $(1 - \beta_e\mu_e^{\prime})\gamma_e$, though the converse is not always true.

The full biased distribution of electron velocities is proportional to
\begin{equation}
    \overbrace{\sigma_\tn{KN}\left((1 - \beta_e\mu_e^{\prime})\gamma_e\varepsilon^{\prime}\right)}^\tn{Cross section bias} \times
    \underbrace{(1-\beta_e\mu_e^{\prime})}_{\mathclap{\tn{Encounter rate bias}}}  \times 
    \overbrace{f_\tn{MJ}(\gamma_e,\Theta)\,d\gamma_ed\mu_e^{\prime}}^\tn{Thermal base population},
\end{equation}
from which \verb|SPIRO| samples electron velocities in a series of rejection steps.

\section{Pair check} \label{app:pair_check}
A photon will Compton scatter on a positron the same way it would on an electron. It is useful to define the pair loading factor $\zeta_\pm \equiv n_\pm/n_p$, where $n_p$ is the number density of protons and $n_\pm$ is the number density of electrons plus positrons (including primary electrons). The Compton opacity of the plasma depends linearly $\zeta_\pm$. The current version of \verb|SPIRO| assumes $\zeta_\pm=1$ (no pairs).

The cross-section for pair production is given by \citep[e.g.,][]{Ito2018}
\begin{equation}
    \sigma_{\gamma\gamma}(\beta_*) = \frac{3\sigma_\tn{T}}{8}(1-\beta_*^2)\left[(3-\beta_*^4)\operatorname{artanh}(\beta_*)-\beta_*(2-\beta_*^2)\right],
\end{equation}
where $\beta_*$ is the dimensionless speed of the created positron (or electron) in the center-of-momentum frame. If we have two photons with dimensionless energies $\varepsilon_1$ and $\varepsilon_2$ in some frame, then the corresponding value of $\beta_*$ will be
\begin{equation}
    \beta_*(\varepsilon_1,\varepsilon_2,\mu_{\gamma\gamma}) = \sqrt{1 - \frac{2}{\varepsilon_1\varepsilon_2(1-\mu_{\gamma\gamma})}},
\end{equation}
where $\mu_{\gamma\gamma}$ is the cosine of the angle between the photons in the given frame. For the interaction to be possible we need 
\begin{equation} \label{eq:pair_prod_threshold}
    \frac{1}{2}\varepsilon_1\varepsilon_2(1-\mu_{\gamma\gamma}) \ge 1,
\end{equation}
to hold. Otherwise the available energy in the center-of-momentum frame will be smaller than $2m_e c^2$ and $\sigma_{\gamma\gamma} = 0$ by default.

For the scenario studied in this paper, the condition in Equation \eqref{eq:pair_prod_threshold} is almost never satisfied for photons inside the same the cell. Looking at Figure \ref{fig:comoving_spectra}, we see that a small fraction of photons reach $\varepsilon^\prime>1$ in the plasma comoving frame inside the RMS's. Encounters between those photons will be rare and almost never head-on ($\mu_{\gamma\gamma}^\prime=-1$). For this reason we do not expect pair production (and related annihilation) to have any significant effect on the radiation field. However, the high photon-to-proton ratio, $\zeta\sim10^5$, means that the Compton opacity of the plasma can be raised significantly even if only a small fraction of the photons turn into pairs. This concern warrants a more careful treatment. 

Part of \verb|SPIRO|'s output are 2D-histograms for the radiation inside each cell at every timestep. These histograms bin the photons based on their energies and directions in the plasma comoving frame. We can directly calculate the corresponding rate density of pair production in a cell using
\begin{equation} \label{eq:pair_prod_rate_density}
    \dot n_{\pm,\tn{prod}} = \sum_{i,j,k,l} w_{ij} w_{kl} \int_0^\pi (1 - \tilde\mu(\mu_j,\mu_l,\phi))c\sigma_{\gamma\gamma}(\beta_*(\varepsilon_i,\varepsilon_k, \tilde\mu(\mu_j,\mu_l,\phi))) \frac{d\phi}{\pi},
\end{equation}
where $\varepsilon_{i}$ is the center of $\varepsilon$-bin $i$, $\mu_{j}$ is the center of $\mu$-bin $j$, $w_{ij}$ is the number of photons in $(\varepsilon,\mu)$-bin $(i,j)$, and
\begin{equation}
    \tilde\mu(\mu_1,\mu_2,\Delta\phi) = \mu_1\mu_2 + \sqrt{1-\mu_1^2}\sqrt{1-\mu_2^2}\cos(\Delta\phi),
\end{equation}
is the expression for the cosine of the angle between two vectors, given the cosines of their respective polar angles $\mu_1$, $\mu_2$, and the difference between their azimuthal angles $\Delta\phi$. The integral averages over the assumed axial symmetry of the radiation and can be computed numerically.\footnote{This integral was evaluated with a 5-point trapezoidal sum when computing the pair checks for the two simulations in this paper.}
 
The mean lifetime for a positron in the plasma is an increasing function of its kinetic energy relative to the comoving frame of the electron gas. But any positron that get produced will be thermalized to the local plasma temperature extremely quickly. The thermally averaged cross section for annihilation is given by the integral \citep{Svensson1982}
\begin{equation}
    \left\langle\sigma_\tn{ann} \beta_\tn{rel}\right\rangle = \frac{\Theta^{-1}}{\sqrt{2}\tn{K}_2(\Theta^{-1})^2}\int_1^\infty \frac{\gamma_\tn{rel}^{2}-1}{\sqrt{\gamma_\tn{rel}+1}}K_{1}\!\left(\frac{\sqrt{2}\sqrt{\gamma_\tn{rel}+1}}{\Theta}\right)\sigma_\tn{ann}\left(\gamma_\tn{rel}\right)d\gamma_\tn{rel},
\end{equation}
where $\tn{K}_1$ and $\tn{K}_2$ are modified Bessel functions of the second kind, $\Theta$ is the dimensionless plasma temperature, and the annihilation cross section is given by
\begin{equation}
    \sigma_\tn{ann}(\gamma_\tn{rel}) = \frac{3\sigma_\tn{T}}{8} \frac{1}{1+\gamma_\tn{rel}}\left(\frac{\gamma_\tn{rel}^2+4\gamma_\tn{rel}+1}{\gamma_\tn{rel}^2-1}\ln\left(\gamma_\tn{rel}+\sqrt{\gamma_\tn{rel}^2-1}\right) - \frac{\gamma_\tn{rel}+3}{\sqrt{\gamma_\tn{rel}^2-1}}\right),
\end{equation}
where $\gamma_\tn{rel}$ is the relative Lorentz factor between the electron and the positron. This integral can be approximated with \citep{Ito2018}
\begin{equation}
    \left\langle\sigma_\tn{ann} \beta_\tn{rel}\right\rangle \approx
    \frac{3\sigma_\tn{T}}{8}\left[1 + \frac{2\Theta^2}{\ln(2 e^{-C_\tn{EM}}\Theta+1.3}\right]^{-1},
\end{equation}
where $C_\tn{EM}$ is the Euler--Mascheroni constant.
The corresponding rate density is
\begin{equation} \label{eq:annihilation_rate_density}
    \dot n_{\pm,\tn{ann}} = -2\left\langle\sigma_\tn{ann} \beta_\tn{rel}\right\rangle c n_{e^+} n_{e^-},
\end{equation}
where the factor of 2 comes from the fact that two particles are destroyed per annihilation event. Such a factor is technically also present in Equation \eqref{eq:pair_prod_rate_density}. But it gets canceled by the symmetry factor due to double counting in the sum. This does not happen in Equation \eqref{eq:annihilation_rate_density} because electrons and positrons are mutually distinguishable as particles.

By using $n_{e^+} n_{e^-}= n_p^2 (\zeta_\pm^2-1)/4$ and setting
\begin{equation}
    \dot n_\pm = \dot n_{\pm,\tn{prod}} + \dot n_{\pm,\tn{ann}} = 0,
\end{equation}
we can obtain the equilibrium value of the pair loading for any fixed value of $\dot n_{\pm,\tn{prod}}$ and $\Theta$. We get
\begin{equation}
    \zeta_\pm^\tn{eq} \approx \sqrt{1 + \frac{16}{3\sigma_\tn{T}n_p^2c} \left[1 + \frac{2\Theta^2}{\ln(2 \Theta e^{-C_\tn{EM}}+1.3}\right] \dot n_{\pm,\tn{prod}}}.
\end{equation}
The actual pair loading $\zeta_\pm$ may differ from the instantaneous value of $\zeta_\pm^\tn{eq}$ since it takes time to settle into equilibrium. But the highest value of $\zeta_\pm^\tn{eq}$ attained during a simulation sets an upper bound for the highest value of $\zeta_\pm$ that could have occurred somewhere in the plasma.

For both simulations in this paper, the value of $\zeta_\pm^\tn{eq}$ stayed below $1+10^{-5}$ in all cells at all times. This implies that pair production had no appreciable effect on the Compton opacity of the plasma.

\section{Propagation details} \label{sec:prop_details}
Kinematic drift in $\mu$ for a freely propagating photon is caused by the fact that its momentum becomes more and more aligned with the local radial direction over time. To find an expression for $\Delta t_\tn{KD}$ which is safe but efficient, we note that 
\begin{equation}
    t_\tn{mf} = \frac{1}{(1 - \beta_r\mu) n_e c \tilde\sigma(\varepsilon^{\prime}, \Theta_\tn{pl})}
\end{equation}
only depends on $\mu$ via $1-\beta_r\mu$ and $\varepsilon^{\prime}=(1-\beta_r\mu)\Gamma\varepsilon$. If $\mu$ is increased from $\mu(t_0)=\mu_0$ to $\mu(t_0+\Delta t_\tn{KD})$, then the relative change $l$ in both of these quantities will be the same:
\begin{equation}
    l = \frac{|(1-\beta_r\mu(t_0+\Delta t_\tn{KD}))\varepsilon\Gamma - (1-\beta_r\mu_0)\varepsilon\Gamma|}{(1-\beta_r\mu_0)\varepsilon\Gamma} = \frac{|(1-\beta_r\mu(t_0+\Delta t_\tn{KD}))-(1-\beta_r\mu_0)|}{1-\beta_r\mu_0}.
\end{equation}
We fix $l$ to be the largest relative change we will tolerate. Using Equation \eqref{eq:mu_of_t} we can solve for $\Delta t_\tn{KD}$, which gives two solutions:
\begin{equation}
    \Delta t_\tn{KD}^\pm = r_0\left(-\mu_0 \pm \frac{\sqrt{1-\mu_0^{2}}}{\sqrt{\left(\mu_0 + \dfrac{1-\beta_r\mu_0}{|\beta_r|}l\right)^{-2}-1}}\right).
\end{equation}
We pick the $\Delta t_\tn{KD}$ to be the smallest positive real solution. If both solutions are negative or non-real, then no amount of future propagation will cause a relative change of $l$. In this case we let $\Delta t_\tn{KD}=+\infty$. The two simulations mentioned in this paper used $l=0.01$.

When calculating the time until the next cell change $\Delta t_\tn{CC}$, we need to be able to model the in-flight motion of the interfaces between the discrete timesteps. The way \verb|SPIRO| staggers the packet propagation with the hydro-steps, the interface positions at the start of the next hydro-step are known before the photon propagation between the current and next hydro-step is computed.\footnote{Any velocity changes due to momentum deposition by the photons will be accounted by the hydrodynamics in the subsequent timestep. The requirements for temporal resolution in Section \ref{sec:numerical_validity} ensures that this method is stable and accurate.}
Let $t=0$ at the start of a particular hydro-step, whose step size is $\delta t$. Consider an interface that is currently located at $r=r_\tn{I}$ and will end up at $r=r_\tn{I,next}$ at $t=\delta t$. Assuming that its in-flight radial velocity is constant, it will be given by $\beta_\tn{I} = (r_\tn{I,next}-r_\tn{I})/\delta t$. The time dependent interface location will then be $r_I + \beta_\tn{I}t$. Equation \eqref{eq:r_of_t} implies the time it takes for a photon packet, located at $r=r_0$ at $t=t_0$, to reach this interface can be found by solving
\begin{equation}
    r_I + \beta_\tn{I}(t_0+\Delta t_\tn{CC}) = \sqrt{r_0^2 + 2 r_0 \mu_0 \Delta t_\tn{CC} +  (\Delta t_\tn{CC})^2}.
\end{equation}
The two candidate solutions are
\begin{equation}\label{eq:catastrophic_cancellation}
    \Delta t_\tn{CC}^\pm = \frac{-B\pm\sqrt{B^2-AC}}{A}
\end{equation}
where
\begin{align}
    A &= 1-\beta_\tn{I}^2, & B &= r_0 \mu_0 - (r_\tn{I}+\beta_\tn{I}t_0)\beta_\tn{I}, & C &= r_0^2-(r_\tn{I}+\beta_\tn{I}t_0)^2.
\end{align}
Negative solutions corresponds to crossings in the past. Non-real solutions are treated as $+\infty$ and imply that crossings are never possible. In the packet propagation algorithm, these solutions are computed for the two interfaces on either side of the cell that the packet is in. The value of $\Delta t_\tn{CC}$ is given by the smallest positive candidate.

When $\beta_\tn{I}$ approaches 1, Equation \eqref{eq:catastrophic_cancellation} becomes numerically unstable for the root closest to zero due to catastrophic cancellation in the numerator. This can be sidestepped by calculating the roots as
\begin{equation}
    \frac{-B-\sgn(B)\sqrt{B^2-AC}}{A} \quad \text{and} \quad \frac{C}{-B-\sgn(B)\sqrt{B^2-AC}}.
\end{equation}


\bibliography{References}{}
\bibliographystyle{aasjournalv7}

\end{document}